\documentclass[12pt]{article}
\pdfoutput=1
\usepackage{multirow}
\usepackage{jheppub}
\usepackage{ifpdf}
\usepackage{array}
\usepackage{booktabs}
 \newcommand{\ra}[1]{\renewcommand{\arraystretch}{#1}}
\usepackage{amssymb}
\usepackage{color}
\usepackage{graphicx,psfrag, subfigure}
\usepackage{amsmath}
\usepackage{hyperref}
\usepackage{bbold}
\usepackage{blkarray}
\usepackage[applemac]{inputenc}
\usepackage{dsfont}
\usepackage{setspace}
\usepackage{soul}
\definecolor{dm}{cmyk}{.20, 0, .30, 0}
\sethlcolor{dm}

\numberwithin{equation}{section}
 \usepackage{relsize}
\usepackage{tabularx}
\newcommand{\sm}{\scalebox{0.25}[1.0]{\( -  \)}}
\newcommand{\six}{\scalebox{1}[1.0]{\( 6  \)}}
\newcommand{\one}{\scalebox{0.65}[1.0]{\( 1  \)}}

\newcommand{\diag}{\text{diag}}

\def\bi{\bar{\imath}}
\def\bj{\bar{\jmath}}

\def\bW{\overline{W}}
\def\bZ{\bar Z}

\def\M{M_{{\rm{pl}}}}

\def\be{\begin{equation}}
\def\ee{\end{equation}}
\def\bea{\begin{eqnarray}}
\def\eea{\end{eqnarray}}
\def\calc{\mathsmaller {\mathfrak  C}}

\def\be{\begin{equation}}
\def\ee{\end{equation}}
\def\bea{\begin{eqnarray}}
\def\eea{\end{eqnarray}}

\newcommand{\resc}[1]{{\varpi }_{\cal Q}(#1)}

\begin{document}

\begin{titlepage}

\setcounter{page}{1} \baselineskip=15.5pt \thispagestyle{empty}

\bigskip\
\begin{center}
{\Large \bf Planckian Axions in String Theory}\\
\vskip 5pt
\vskip 15pt
\end{center}
\vspace{0.5cm}
\begin{center}
{
Thomas C. Bachlechner, Cody Long, and Liam McAllister}
\end{center}\vspace{0.05cm}

\begin{center}
\vskip 4pt
\textsl{Department of Physics, Cornell University, Ithaca, NY 14853 USA}

\end{center}
{\small  \noindent  \\[0.2cm]
\noindent

We argue that super-Planckian diameters of axion fundamental domains can naturally arise in Calabi-Yau compactifications of string theory.
In a theory with $N$ axions $\theta^i$, the fundamental domain is a polytope defined by the periodicities of the axions, via constraints of the form $-\pi<Q^{i}_{~j} \theta^j<\pi$.
We compute the diameter of the fundamental domain in terms of the eigenvalues $f_1^2\le\ldots\le f_N^2$ of the metric on field space, and also, crucially, the largest eigenvalue of $({ Q}{ Q}^{\top})^{-1}$.  At large $N$, ${ Q}{ Q}^{\top}$ approaches a Wishart matrix, due to universality, and we show that the diameter is at least $N f_{N}$, exceeding the naive Pythagorean range by a factor $>\sqrt{N}$.  This result is robust in the presence of $ P > N$ constraints, while for $P=N$ the diameter is further enhanced
by eigenvector delocalization to $N^{3/2}f_N$.
We directly verify our results in explicit Calabi-Yau compactifications of type IIB string theory. In the classic example with $h^{1,1}=51$ where parametrically controlled moduli stabilization was demonstrated by Denef et al.~in \cite{Denef:2005mm}, the largest metric eigenvalue obeys $f_N \approx 0.013 \M$.  The random matrix analysis then predicts, and we exhibit, axion diameters $>\M$ for the precise vacuum parameters found in \cite{Denef:2005mm}.
Our results provide a framework for achieving large-field axion inflation
in well-understood flux vacua.
}

\vspace{0.3cm}

\vspace{0.6cm}

\vfil
\begin{flushleft}
\small \today
\end{flushleft}
\end{titlepage}
\tableofcontents
\newpage
\section{Introduction}\label{intro}

An important class of inflationary models are those involving super-Planckian displacements of the inflaton field.   These `large-field' scenarios yield a detectably-large spectrum of primordial gravitational wave fluctuations, and can therefore be tested in the coming generation of CMB polarization experiments.   The  predictions of large-field models depend sensitively on the  couplings of the inflaton $\phi$  to the degrees of freedom comprising the ultraviolet completion of gravity (see \cite{Liamsbook} for a review).   As a result, to formulate a large-field model one must make explicit or implicit assumptions about quantum gravity.

A leading proposal for  controlling the ultraviolet sensitivity of large-field inflation is to  incorporate a weakly broken shift symmetry, $\phi \to \phi + \text{const.}$,  in order to protect the  inflaton potential  over a super-Planckian range.  From the viewpoint of the low-energy effective field theory for $\phi$, the shift symmetry is an internally-consistent assumption that renders small renormalizable couplings of $\phi$  radiatively stable.   However, general reasoning about the absence of exact continuous global symmetries in quantum gravity, and specific results from string theory, strongly suggest that not every shift-symmetric effective field theory  coupled to  gravity admits an ultraviolet completion.   To provide a microphysical foundation  for large-field inflation, one must therefore establish the existence of a suitable symmetry in a computable regime of quantum gravity.

A well-motivated strategy is to take the inflaton field(s) to correspond to one or more axions in a compactification of string theory.   Axions are numerous in Calabi-Yau compactifications, and in the absence of  specific fluxes and branes that introduce monodromy, the potential for each axion vanishes to all orders in perturbation theory.   The leading potential then arises from nonperturbative effects, and is sinusoidal.   For a single  dimensionless axion $\theta$,  the Lagrangian takes the form
\begin{equation}
{\mathcal L}=\frac{1}{2}f^2 (\partial\theta)^2 - \Lambda^{4}\left(1-\cos(\theta)\right)\,,  \label{equ:natural}
\end{equation}  where $\Lambda$ is a dynamically-generated scale, and the parameter $f$ is known as the axion decay constant.   The canonically-normalized field with mass dimension one is then $\phi \equiv f\theta$.  In vacua of string theory involving small numbers of axions, the axion decay constants are typically small, $f\ll \M$, in the regime of weak coupling and large volume where perturbative computation of the effective action is valid \cite{Banks:2003sx} (see also \cite{Svrcek:2006yi}).   The fundamental domain for $\phi$ has diameter $2\pi f$, and as a result (\ref{equ:natural}) does not give rise to a realistic inflationary model in the absence of monodromy.

The purpose of this paper is to compute the diameter of the fundamental domain in an extension of (\ref{equ:natural}) to a totally general system with $N \gg 1$ axions, keeping track of all factors parametric in $N$.  This may sound straightforward, but influential early works \cite{KNP,Nflation}  as well as more recent analyses \cite{Czerny:2014qqa,Czerny:2014xja,Tye:2014tja,Kappl:2014lra,Ben-Dayan:2014zsa,Long:2014dta,Ben-Dayan:2014lca,Bachlechner:2014hsa,Ali:2014mra} --- including our own works on the subject ---  captured only fragments of the full field range that is present in {\it{generic}} large $N$ systems, including explicit string compactifications.   In this work we unify the field range enhancements arising in N-flation \cite{Nflation}, including kinetic alignment from eigenvector delocalization \cite{Bachlechner:2014hsa},  with the full field range arising from the decay constant alignment mechanism of Kim, Nilles, and Peloso \cite{KNP}.   We then argue that enhancements of the field range by a factor $\sim N$  compared to the naive expectation are automatically present in a broad class of theories.   Finally, we illustrate our results in a completely explicit compactification of type IIB string theory.

The organization of this paper is as follows. In \S\ref{sec:domain} we review the definition of the fundamental domain in a system of $N$ axions and give an intuitive estimate for its diameter, along with an overview of our results. In \S\ref{sec:align} we compute the diameter of the fundamental domain and  describe the mechanism of kinetic alignment.
Dynamic alignment is described in \S\ref{sec:infdynamics}.  In \S\ref{sec:sugra} we discuss the embedding of axions in supergravity and estimate the leading coupling to saxions in a supersymmetric vacuum. Then, in \S\ref{sec:DDFGK} we analyze the diameter of the fundamental domain in the F-theory compactification described in \cite{Denef:2005mm}.
In \S\ref{unifiedtheory} we explain how our approach gives a unified picture for the geometry of axion field spaces.
We conclude in \S\ref{sec:conclusion}. In appendix \ref{app:RMT} we briefly review a few facts about random matrix theory that are  needed in this work. In appendix \ref{app:toric} we  give examples of nontrivial fundamental domains arising in string compactifications on Calabi-Yau hypersurfaces in toric varieties.

\section{The Diameter of the Fundamental Domain}\label{sec:domain}

Consider a theory of $N$  axions $\theta^i$ that at the perturbative level enjoy the continuous shift symmetries $\theta^i \to \theta^i + \text{const.}$, so that
a general two-derivative action for the $\theta^i$ can be written
\be
{\mathcal L}={1\over 2}K_{ij} \partial\theta^i\partial\theta^j\,,
\ee
where $K_{ij}$ is a metric\footnote{In a supersymmetric theory, $K_{ij}$ arises from the K\"ahler metric on field space, but our arguments apply with or without supersymmetry.} on the field
space ${\cal M}$, which is diffeomorphic to $\mathbb{R}^N$.

Nonperturbative contributions from instantons give rise to a potential that is a sum of sinusoidal terms,
\be
{\mathcal L}={1\over 2}K_{ij} \partial\theta^i\partial\theta^j - \sum_{i=1}^N \Lambda_i^4 \left[1-\cos\left({\cal Q}^i_{\, j}\theta^j \right) \right]\,,
\ee where ${\cal Q}$ is an $N \times N$  matrix with integer entries.\footnote{Throughout this work we will assume that the  number $P$ of nonperturbative terms is at least $N$; that is, all axions are stabilized.   In the present discussion we take $P=N$ for simplicity, describing the case $P\ge N$ in \S\ref{sec:align}.}
This potential breaks the continuous shift symmetries to discrete shifts. The associated periodic identifications,
\begin{equation}
\Gamma_i: \quad{\cal Q}^i_{\, j}\theta^j \cong {\cal Q}^i_{\, j}\theta^j + 2\pi \,,
\end{equation}
define $N$ pairs of identified hyperplanes in $\mathbb{R}^N$.
By the fundamental domain,  we mean  the intersection of all the identifications, ${\cal M}_{ \Gamma}\equiv{\cal M}/\Gamma_1\cap \dots\cap{\cal M}/\Gamma_N \subset \mathbb{R}^N$, i.e.~the region inside all pairs of hyperplanes.  For the problem of large-field inflation, an interesting invariant quantity is the diameter of ${\cal M}_{\Gamma}$, measured in units where $\M=1$ (which we shall use for the remainder).
This diameter, which we will denote by ${\cal D}$, corresponds to the  magnitude of the maximal rectilinear displacement that the canonical field ${\boldsymbol \Phi}$  can undergo (in the absence of monodromy, which would allow traversing multiple copies of ${\cal M}_{\Gamma}$, as in \cite{SW,MSW}.)   As such, ${\cal D}$ is a proxy for the field range relevant for large-field axion inflation.
Clearly, ${\cal D}$ depends on the identifications $\Gamma_i$: the fundamental domain is bounded by adjacent maxima of each of the sinusoidal terms.

To compute ${\cal D}$, it is convenient to first perform the $GL(N,\mathbb{R})$  transformation
\begin{equation}
{\boldsymbol \phi}={\mathbf Q}\, {\boldsymbol \theta} \label{phitheta}\,.
\end{equation}
In the $\phi^i$ basis, the hyperplanes defining the identifications are orthogonal, and form the faces of an $N$-cube of side $2\pi$.
The kinetic matrix is then given by
\be
{\mathbf \Xi}= ({\mathbf Q}^{-1})^\top {\mathbf K}\, {\mathbf Q}^{-1}\,,
\ee and the Lagrangian takes the form
\be\label{eqn:philag}
{\mathcal L}={1\over 2}\partial{\boldsymbol \phi}^\top {\mathbf \Xi}\, \partial{\boldsymbol \phi}-\sum_{i=1}^N \Lambda_i^4 \left[1-\cos\left(\phi_i \right) \right]\,,
\ee
At the perturbative level, the metric on field space is independent of the axions, so ${\mathbf \Xi}$ is a constant matrix, up to nonperturbatively small corrections.
However, ${\mathbf \Xi}$ is in general not diagonal in the $\phi^i$ basis.   Thus, the ${\boldsymbol \phi}$ are related to the canonically-normalized fields ${\boldsymbol \Phi}$ by a further $GL(N,\mathbb{R})$  transformation  (i.e., a diagonalization of ${\mathbf \Xi}$ by an orthogonal transformation, combined with a rescaling by the eigenvalues $\xi_i^2$ of ${\mathbf \Xi}$).

We should stress the elementary but crucial point that writing ${\mathbf Q}= {\mathbb 1}$ in the $\theta^i$ basis is {\it{not}} equivalent to beginning with a theory for which ${\mathbf Q}\ne {\mathbb 1}$ in the $\theta^i$ basis, and then performing the transformation (\ref{phitheta}) that renders ${\cal M}_{\Gamma}$ hypercubic.    In the former case, the metric on ${\cal M}_{\Gamma}$ is ${\mathbf K}$, while in the latter case it is ${\mathbf \Xi}= ({\mathbf Q}^{-1})^\top \mathbf K\, {\mathbf Q^{-1}}$.  Because ${\mathbf Q}$ is generally not orthogonal, the eigenvalues of ${\mathbf \Xi}$ differ from those of ${\mathbf K}$.\footnote{The fact that the axion field range is large when the smallest eigenvalue of $\mathbf Q^\top \mathbf Q$ is small is the core of the Kim-Nilles-Peloso (KNP) mechanism for decay constant alignment \cite{KNP}, which was generalized to the case $N>2$ by Choi, Kim, and Yun in \cite{Choi:2014rja} and explored by Higaki and Takahashi in \cite{HT1,HT2}.
See the discussion in \S\ref{unifiedtheory}.}

To summarize, the task is to determine the invariant diameter ${\cal D}$ of the fundamental domain ${\cal M}_{\Gamma}$.   To do so, one must specify the identifications $\Gamma$, but these are {\it{not}} invariant under changes of coordinates: there is a preferred `lattice' basis $\phi^i$ in which the periodic identifications are defined by the faces of a hypercube of side length equal to (say) $2\pi$.  This matters, because $GL(N,\mathbb{R})$  transformations in systems with $N \gg 1$  can readily change the eigenvalues of matrices --- including the kinetic matrix, as we shall show ---  by factors that are parametric in $N$.   One must therefore be careful to specify the metric ${\mathbf K}$ and the identifications $\Gamma$ in the same basis, and then proceed to compute the invariant distance ${\cal D}$.

Thus far all of our statements have been deterministic, and amount to saying that  in a theory that specifies $\mathbf Q\neq \mathbb{1}$ and $\mathbf K$, it is obviously incorrect to take $\mathbf Q= \mathbb{1}$ when computing ${\cal D}$.  We now turn to making statistical arguments, based on the phenomena of universality, eigenvalue repulsion, and eigenvector delocalization in random matrix theory.  We will argue that ${\cal D} \gg {\cal D}|_{\mathbf Q=\mathbb{1}}$, with an enhancement that is parametric in $N$.   The precise degree of enhancement depends on the forms of $\mathbf Q$ and $\mathbf K$, as we will explain below.

Because the argument involves a number of independent computations of the behavior of $N\times N$ matrices at large $N$, here we will give an accessible overview of  main steps of the calculation.
The complete calculation follows in \S\ref{sec:align}, while background on relevant results from  random matrix theory appears in appendix \ref{app:RMT}.

To compute ${\cal D}$, it is convenient to work in the $\phi^i$ basis, where ${\cal M}_{\Gamma}$ is an $N$-cube of side $2\pi$.
Hypersurfaces in ${\cal M}$ of constant invariant distance $r$ from the origin are ellipsoids ${\cal E}_r$ defined by
\begin{equation}
{\boldsymbol \phi}^\top {\mathbf \Xi}\, {\boldsymbol \phi} = r^2\,.
\end{equation}
The diameter ${\cal D}$ is then given by ${\cal D}=2r_{\rm{max}}$, where $r_{\rm{max}}$ is the largest value of $r$ for which ${\cal E}_r$ intersects ${\cal M}_{\Gamma}$.

The largest possible ${\cal D}$ arises if the shortest principal axis of ${\cal E}_r$, corresponding to the eigenvector $\mathbf \Psi_N^{\Xi}$ of $\mathbf \Xi$ with the largest eigenvalue $\xi_N^2$, is parallel to a diagonal of the $N$-cube.  In that case we have
\begin{equation}\label{diamestimate}
{\cal D}_{\rm{max}} =2\pi \xi_N \sqrt{N}\,.
\end{equation}
In general, $\mathbf \Psi_N^{\Xi}$ will not point precisely along a diagonal, but due to vast number of diagonals in a hypercube, $\mathbf \Psi_N^{\Xi}$ is with high probability very nearly parallel to a diagonal, so that (\ref{diamestimate}) is an accurate estimate.

In order to estimate the typical diameter, we first assume that the metric on field space is trivial, $\mathbf K=f^2 {\mathbb 1}$, and that the matrix ${\mathbf Q}$ is sparse and contains random integers with r.m.s.~size $\sigma_{Q}$.   Even though ${\mathbf Q}$ is sparse, when a fraction $\gtrsim 2/N$ of its entries are non-vanishing the matrix $\mathbf Q^\top \mathbf Q$ approaches its universal limit of a random matrix in the Wishart ensemble. In this random matrix ensemble, strong eigenvalue repulsion forces the smallest eigenvalue $\lambda_1$ to obey $\lambda_1 \lesssim \sigma_{Q}^2/N$. If the non-vanishing entries of ${\mathbf Q}$ have scale ${\mathcal O}(1)$, the minimum scale of the matrix ${\mathbf Q}$ is given by $\sigma_{Q}\approx {2/ \sqrt{N}}$. In this case, from (\ref{diamestimate}) we find that
\be
{\cal D}\gtrsim  N^{3/2}f\,.
\ee
In \S\ref{sec:align} we extend this logic to cases in which the metric is either a Wishart matrix or a heavy-tailed matrix, as well as to the general case where the number of constraints, $P$, exceeds the number of axions, so that ${\cal Q}$ is rectangular.
Furthermore, we will show in \S\ref{sec:infdynamics} that the lightest canonical field is generically aligned with the largest diameter of the fundamental domain.

\section{Kinetic Alignment}\label{sec:align}

In the previous section we outlined our strategy for determining the diameter of ${\cal M}_{\Gamma}$. We now turn to a more detailed analysis and derive the main results of this work.

Let us consider an action for $N$ axions whose potential is generated nonperturbatively, and is periodic in the axions.  This action will be further motivated in \S\ref{sec:sugra}, when we discuss embeddings of our results in supergravity theories that arise as effective theories in string compactification.   We assume that there are $P \ge N$ nonperturbative terms in the potential,  so that the most general Lagrangian for the axions $\boldsymbol \theta$ is given by
\be
{\mathcal L}={1\over 2}K_{ij} \partial\theta^i\partial\theta^j-\sum_{i=1}^P \Lambda_i^4 \left[1-\cos\left(\mathcal Q^i_{~j}\theta^j \right) \right]\,,
\ee
where we chose units such that each of the axions has the shift symmetry $\mathcal Q^i_{~j}\theta^j \rightarrow \mathcal Q^i_{~j}\theta^j+2\pi$, and the entries of the $P\times N$ matrix
${\cal Q}$ are integers.
Without loss of generality we can decompose $\mathcal Q$ as
\be
{\mathcal Q}= \left(\begin{tabular}{c} $\mathbf Q$\\ $\mathbf Q_\text{R}$\end{tabular}\right)\, ,
\ee
where $\mathbf Q$ is a square, full rank matrix and ${\mathbf Q}_\text{R}$ is a rectangular $(P-N)\times N$ matrix. Now, define fields $\boldsymbol \phi$ as
\be
\boldsymbol \phi= \mathbf Q \, \boldsymbol \theta\, ,
\ee
such that
\be
\mathbf Q \, \boldsymbol\theta=\left(\begin{tabular}{c} ${\mathbb 1}$\\ $\mathbf Q_\text{R} \mathbf Q^{-1}$\end{tabular}\right)\boldsymbol\phi\,.
\ee
Here we are making a field redefinition to simplify $N$ terms in the potential, while $P-N$ terms will depend on linear combinations of the $\phi^i$. Therefore, the fundamental domain is given by an $N$-cube of side length $2\pi$, cut by $2(P-N)$ hyperplanes that constitute the remaining constraints:
\be\label{eqn:constraints}
-\pi\le \left( \mathbf Q_\text{R} \mathbf Q^{-1}\boldsymbol \phi \right)^i \le \pi \, \quad \forall \, i\, .
\ee
Some comments are in order. If the matrix $\mathcal Q$ were square, then this field redefinition would be unique, and would uniquely define what we mean by an axion: a field that appears in the potential as the argument of a cosine. In the rectangular case there are more cosines than fields, so the definition of an axion is not physical, but depends on a choice of basis. However, the diameter of the fundamental domain is physical and basis-independent. In terms of the axions $\phi^i$ the Lagrangian becomes
\be\label{eqn:lag}
{\mathcal L}={1\over 2}\partial{\boldsymbol \phi}^\top {\mathbf \Xi}\, \partial{\boldsymbol \phi}-\sum_{i=1}^N \Lambda_i^4 \left[1-\cos\left(\phi^i \right) \right]-\sum_{i=1}^{P-N} \Lambda_i^4 \left[1-\cos\left(\left(\mathbf Q_\text{R} \mathbf Q^{-1} \boldsymbol \phi\right)^i \right) \right]\,,
\ee
where, as before,
\be
\mathbf \Xi= (\mathbf Q^{-1})^\top \mathbf K \, \mathbf Q^{-1}\,
\ee
is the kinetic matrix of our choice of axions $\phi^i$, with eigenvalues $\xi_i^2$. So far, we have performed a field redefinition so that the fields $\boldsymbol \phi$ appear as the arguments of $N$ of the cosines.
Finally, the canonically normalized fields are given by
\be\label{canfields}
\boldsymbol \Phi=\diag(\xi_i)\, \mathbf S^\top_{\Xi} \, \boldsymbol\phi\,,
\ee
where $\mathbf S^\top_{\Xi}$ diagonalizes $\mathbf \Xi$,
\begin{equation}
\mathbf S^\top_{\Xi}\, \mathbf \Xi\, \mathbf S^{\phantom{T}}_{\Xi} = \diag(\xi_i^2)\,.
\end{equation}
We will use (\ref{canfields}) in order to determine canonically normalized distances on moduli space.

In general, no closed form expression is available for the maximal diameter of the polytope defining the fundamental domain ${\cal M}_{\Gamma}$.
Instead, to obtain a lower bound on the maximal diameter, we will compute the diameter ${\cal D}$ of ${\cal M}_{\Gamma}$ along the direction of a particular unit vector $\hat{\mathbf v}$ in the $\boldsymbol \phi$ basis.
A useful choice is to take $\hat{\mathbf v}$ to be the direction defined by a linear superposition of kinetic matrix eigenvectors $\Psi^{\Xi}_i$, weighted in proportion to the  square roots $\xi_i$ of the corresponding eigenvalues $\xi_i^2$:
\be\label{vvector}
\mathbf{v}= \sum\limits_i \xi_i  \boldsymbol \Psi^{\Xi}_i\, .
\ee
We now define an operator $\resc{\mathbf w}$ that rescales a vector $\mathbf w$ to saturate the constraint equations~(\ref{eqn:constraints}) of the fundamental domain:
\be\label{rescale}
\resc{\mathbf w}\equiv {2\pi\over \text{Max}_i\left( \{|({\cal  Q} Q_R^{-1} {\mathbf w})_i|\}\right) }\times {\mathbf w}\,.
\ee
In the geometric picture of \S\ref{sec:domain}, $\mathbf w$ ends on an ellipsoid ${\cal E}_w$ at invariant distance $r_w$ from the origin, and (\ref{rescale})  rescales $\mathbf w \to \resc{\mathbf w}$ so that ${\cal E}_{\resc{\mathbf w}}$ just intersects ${\cal M}_{\Gamma}$.

Using the rescaling operator and (\ref{canfields}), we find that the canonically normalized diameter of the fundamental domain along the direction $\hat{\mathbf v}$ is
\be\label{kinrange}
{\cal D}=\left \lVert \diag{\xi_i}\, \mathbf S_{\Xi}^\top \resc{\hat{\mathbf v}}\right \rVert=\left \lVert \resc{\hat{\mathbf v}} \right\rVert \sqrt{\sum_{i=1}^N \xi_i^4}\,,
\ee
where we used that the eigenvectors are orthonormal: $ \sum_i \mathbf S_{\Xi}^\top  \xi_i  \boldsymbol \Psi^{\Xi}_i=\mathbf S_{\Xi}^\top\, \mathbf S^{\phantom{T}}_{\Xi}  \, \boldsymbol \xi=\boldsymbol\xi$.   As a check, in the special case of $\mathcal Q={\mathbb 1}$ and $\mathbf K=f^2{\mathbb 1}$, we can evaluate (\ref{kinrange}) analytically and obtain the familiar N-flation result:  ${\cal D}=2\pi\sqrt{N}f$.

While (\ref{kinrange}) gives an analytic expression for the diameter of the fundamental domain along an arbitrary direction, it is only useful once the periodicities and the kinetic matrix are defined. We now turn to evaluating the diameter of a {\it generic} fundamental domain. To that end, we assume that the integer entries of the matrix $\cal Q$ are independent and identically distributed (i.i.d.). For a sufficiently large number of non-vanishing entries, the matrix ${\cal Q}^\top {\cal Q}$ then approaches its universal limit of a Wishart distribution\footnote{See also appendix \ref{app:RMT} for a brief review of basic facts from random matrix theory.} \cite{2006math.ph...3038D, 2007JPhA...40.4317V, 2316552, TaoVu, Erdos}. In particular, assuming the entries of $\cal Q$ are of similar scale, the universal limit is reached when a fraction $\gtrsim 2/N$ of the entries in $\cal Q$ are non-vanishing. In the following, we will assume that the universal limit has been reached and $\cal Q$ consists of random integers of similar scale. We will consider three different models for the metric on field space: the identity matrix, a Wishart matrix, and a heavy-tailed matrix.

The above assumptions are motivated by compactifications of type IIB string theory, as we discuss in \S\ref{sec:DDFGK}.
Furthermore, metrics of Wishart and heavy-tailed type are compelling models for metrics on K\"ahler moduli spaces \cite{Long:2014fba}.

\subsection{Diameter estimates}
In order to evaluate the diameter of the fundamental domain (\ref{kinrange}), we need an estimate for the quantity $\lVert \resc{\hat{\mathbf v}}\rVert$ that corresponds to the dimensionless diameter in the direction $\hat{\mathbf v}$.
In general, we can compute the diameter directly from the entries of the matrix ${\cal Q}$ and the metric $\mathbf K$. In order to obtain the typical diameter for a generic matrix $\cal Q$, we assume that its integer entries are i.i.d.~random variables.  The scale of the matrix ${\cal Q}$ is set by $\sigma_{\cal Q}=\langle {\cal Q}\rangle_{\text{r.m.s.}}$.
In the resulting ensemble of kinetic matrices $\mathbf \Xi$,  which is approximately rotationally invariant,
the eigenvectors $\Psi^{\boldsymbol\Xi}_i$ are uniformly distributed on the unit sphere, so that
the unit vector (\ref{vvector}) has normally distributed entries with standard deviation $1/\sqrt{N}$,
\be
\hat{v}^i\in {\mathcal N}(0,1/\sqrt{N})\,.
\ee
This phenomenon is known as eigenvector delocalization.
The median size of the largest entry evaluates to
\be
\text{Max}(\{| \hat{v}_{i}|\})={\sqrt{2} \,\text{erf}^{-1}(2^{-1/N})\over \sqrt{N}} \equiv {\ell_N \over \sqrt{N}}\,.
\ee
For the case of a square matrix ${\cal Q}= \mathbf Q$, the constraints for the fundamental domain simply become $\text{Max}|v_i|\le \pi$ and therefore (\ref{rescale}) immediately becomes
\be\label{squarediam}
\lVert \resc{\hat{\mathbf v}}\rVert=\left\lVert  {2\pi\over \text{Max}_i\left( \{|\hat{v}_i|\}\right) }\times \hat{\mathbf v} \right\rVert={2\pi\over \ell_N}\sqrt{N}\,.
\ee
The result in (\ref{squarediam}) can be understood intuitively from the fact that a  high-dimensional hypercube has vastly more diagonal directions than faces, and therefore a randomly-selected direction is nearly aligned with a diagonal direction, giving a diameter enhanced by $\sqrt{N}$.

For the case where the number of constraints $P$ is larger than the number of axions, ${\cal Q}$ is rectangular. The first $N$ constraints are again $\text{Max}|v_i|\le \pi$, while the remaining constraints are given by $\text{Max}(\{|\mathbf Q_\text{R} \mathbf Q^{-1} {\hat{\mathbf v}}|_i\})\le \pi$. By extensive numerical simulation we observe that the entries of $\mathbf Q_\text{R} \mathbf Q^{-1}{\hat{\mathbf v}}$ for fixed $\mathbf Q$ are Gaussian distributed and the typical standard deviation is given by $\sqrt{2}$, independent of $\sigma_{\cal Q}$, $N$, and $P$. Therefore, the typical size of the largest-magnitude entry of the vector $\mathbf Q_\text{R} \mathbf Q^{-1}{\hat {\mathbf v}} $ is given by
\be
\text{Max}(\{|(\mathbf Q_\text{R}\mathbf Q^{-1}\mathbf v )_i|\})\approx 2\,\text{erf}^{-1}(2^{-{1\over P-N}})\approx \sqrt{4 \log(P-N)}\equiv \l_{P-N} \,.
\ee
These entries are typically much larger than the entries of ${\mathbf v}$, so whenever the number of constraints is larger than the number of axions, the diameter is limited by the additional constraints. The typical diameter then is given by
\be\label{qrectangulardiam}
\lVert \resc{\mathbf v}_\phi\rVert=\left\lVert  {2\pi\over \text{Max}_i\left( \{|v_i|\}\right) }\times \hat{\mathbf v} \right\rVert={2\pi\over \l_{P-N}}\,,
\ee
and the enhancement of the diameter originating from the presence of diagonals is lost.

\begin{figure}
  \centering
  \includegraphics[width=.5\textwidth]{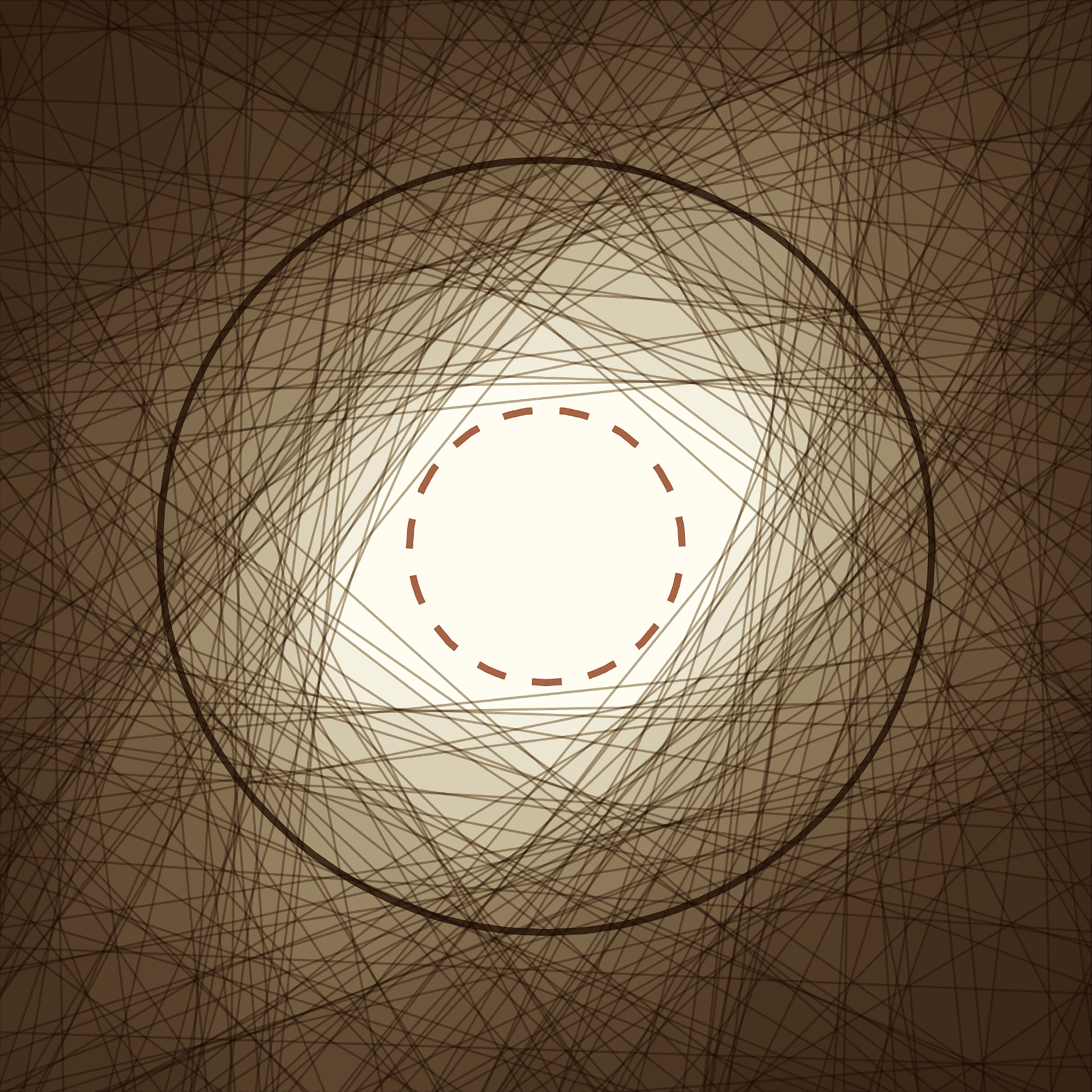}
  \caption{\small  The fundamental domain in the presence of $P \gg N$  constraints, for $N=2$.  The square shown is the domain $|v_{1,2}|\le \pi$, and the lines are 100 hyperplanes defined by $|({\cal Q}Q^{-1}{\mathbf v})_{1,2}|=\pi$, where the elements of ${\cal Q}Q^{-1}$ are Gaussian distributed with standard deviation $\sqrt{2}$. The black circle illustrates the typical location of hyperplanes, while the dashed, red circle illustrates the analytic estimate (\ref{qrectangulardiam}) for the size of the fundamental domain.\label{qfigure}}
  \end{figure}

The loss of enhancement from the presence of diagonals can be understood geometrically,  as illustrated in Figure \ref{qfigure}. The addition of a large number of constraints is defined in terms of $P-N$ hyperplanes, typically located a distance $1/(\sqrt{2N}\sigma_Q)$ from the origin and with normal vectors uniformly distributed on the sphere. The resulting fundamental domain is described by an approximately spherical region around the origin, of diameter $2\pi/\sqrt{2}$.

\subsection{Unit metric}
We now proceed to evaluate the diameter of the fundamental domain in physical units, first assuming the metric to be the identity matrix $\mathbf K=f^2 {\mathbb 1}$. Then, we have for the kinetic matrix $\mathbf \Xi$:
\be \label{koneqtq}
\mathbf \Xi=f^2(\mathbf Q \mathbf Q^\top)^{-1}=f^2 \mathbf S^{\phantom{T}}_{Q^\top Q} \,\diag{(Q_i^{-2})}\, \mathbf S^\top_{Q^\top Q}\,,
\ee
where
\begin{equation}
\mathbf S^\top_{Q^\top Q}  \, \mathbf Q^\top \mathbf Q \, \mathbf S^{\phantom{T}}_{Q^\top Q} = \diag(Q_i^2) \,.
\end{equation}
In the second equality  in (\ref{koneqtq}) we have used the fact that eigenvectors do not change upon inversion. Therefore, we can use eigenvector delocalization of the Wishart ensemble. Considering the diameter in the direction $\boldsymbol \Psi^\Xi_N$, from   (\ref{kinrange}) we obtain the conservative bound
\be\label{kinrange2}
{\cal D}\ge \xi_N\lVert \resc{\boldsymbol \Psi}^{ Q^\top  Q}_N\rVert\,.
\ee
The largest eigenvalue of $\mathbf \Xi$, $\xi^2_N$, obeys (see  appendix \ref{app:RMT})
\be
\xi_N = f {1\over \text{Min}(Q_i)}\,.
\ee
In the large $N$ limit, the median size of the smallest eigenvalue of the Wishart matrix $\mathbf Q^\top \mathbf Q$ is given by $Q_1^2=\calc \sigma_{\cal Q}^2/N$, where $\calc\approx 0.3$ (cf.~(\ref{cmedian})). Finally, the field range is given by
\bea\label{eqn:unitdiam}
{\cal D}\approx f{\sqrt{N} \over\sqrt{\calc} \sigma_{\cal Q}} \lVert \resc{\boldsymbol \Psi^{ Q^\top  Q}_N}\rVert\lesssim \begin{cases}f N^{3/2}~~~&\text{for~}~~~P=N\\ f N~~~&\text{for~}~~~P>N\end{cases}  \,,
\eea
where we used (\ref{minevuniversal}) in the last inequality to set $\sigma_{ \cal Q}^{-1}\lesssim N$.
\subsection{Wishart metric}

To consider a more general metric, let $\mathbf K$ be a Wishart matrix that is diagonalized by $\mathbf S_{K}$ and has  maximum eigenvalue $f_N^2$. The kinetic matrix $\mathbf \Xi$ is then given by
\be\label{wisxi}
\mathbf \Xi=\mathfrak Q^\top \diag{f_i^2} \mathfrak Q\,,
\ee
where $\mathfrak Q= \mathbf S_{K}(\mathbf Q^{-1}) $. While (\ref{wisxi}) is not an inverse Wishart matrix, a reasonable guess for the kinetic matrix is to approximate it as a {\it rescaled} inverse Wishart matrix,
\be
\mathbf \Xi=\mathfrak Q^\top \diag(f_i^2) \, \mathfrak Q\sim \sigma_{f_i^2}(\mathbf Q^\top \mathbf Q)^{-1}\, ,
\ee
where the scale of the eigenvalues is given by $\sigma_{f_i^2}=\langle f_i^2 \rangle_{\text{r.m.s.}}\approx f_{N}^2/4$.
Therefore, $\mathbf \Xi^{-1}$ is approximately an inverse Wishart matrix of scale $\sigma_{\Xi^{-1}}=\sigma_{\cal Q}/\sqrt{\sigma_{f_i^2}}$, and the typical largest eigenvalue of $\mathbf \Xi$ is given by
\be\label{wismaxev}
\xi_N={1\over{\sqrt{{\sigma_{\cal Q}^2 \over\sigma_{f_i^2}}{\calc\over N}}}}={\sqrt{N/\calc}\over 2\sigma_{\cal Q}}f_{N}\,,
\ee
where we used again that $\sigma_{f_i^2}\approx f_{N}^2/4$. The physical field range is then given by
\bea\label{wisrange}
{\cal D}\approx {f_{N}\over 2} {\sqrt{N} \over\sqrt{\calc} \sigma_{\cal Q}} \lVert \resc{\Psi^{Q^\top Q}_N}\rVert\lesssim \begin{cases}f_N N^{3/2}~~~&\text{for~}~~~P=N\\ f_{N} N~~~&\text{for~}~~~P>N\end{cases}  \,,
\eea
We have verified this result through extensive simulations.

\subsection{Heavy-tailed metric}
For a heavy-tailed metric $K$, the eigenvalues $f_i^2$ are distributed with a polynomial fluctuation probability, so the scale $\sigma_{f_i^2}$ is not defined. 
While there are many distinct ensembles of matrices exhibiting heavy tails, a simple model that we will adopt is one where one of the metric entries dominates over all others, so the metric takes the schematic form
\be
K_{11}=f_{N}^2,~K_{ij}\ll f_{N}^2 ~\forall ~i\ne 1 ~\text{or}~ j\ne 1\,.
\ee
See \cite{Long:2014fba} for  examples of heavy-tailed K\"ahler metrics in  explicit string compactifications.

Thus, for the matrix $\mathbf \Xi= (\mathbf Q^{-1})^\top \mathbf K \mathbf Q^{-1} $, $Q^{-1}_{1j}/\lVert Q^{-1}_{1j}\rVert$ is a unit eigenvector corresponding to the eigenvalue $f_{N}^2 \lVert Q^{-1}_{1j}\rVert^2$, while all other eigenvalues are much smaller. The matrix $\mathbf Q$ has entries of scale $\sigma_{\cal Q}$ and is otherwise random, so that the elements of the inverse matrix obey
\be\label{matrixone}
\sum_{i}(Q^{-1})_{1i} Q_{i1}=1\, ,
\ee
where we can approximate the entries of $\mathbf Q$ as Gaussian random variables with vanishing mean and standard deviation $\sigma_{\cal Q}$. The entries of the matrix $\mathbf Q^{-1}$ are then approximately distributed according to the
inverse Gaussian distribution with standard deviation $\sqrt{N}$ in order to satisfy (\ref{matrixone}).  It is then plausible that the sum $ \sigma^2_{Q} \lVert Q^{-1}_{1j}\rVert^2=\sigma_{\cal Q}^2\sum_{i}(Q^{-1})_{1i}^2$ is inverse chi-squared distributed with unit standard deviation:
\be\label{assertion}
 \sigma^2_{Q} |\tilde{Q}^{-1}_{1j}|^2= \sigma^2_{Q} \sum_{i=1}^N (Q^{-1})_{1i}^2\in \chi^{-1}(1)\, .
\ee
While we will not prove this relation, we have verified (\ref{assertion}) numerically, finding an excellent match.
The median of $\lVert Q^{-1}_{1j}\rVert^2$ is then given by
\be
\tilde{\lambda}=\left({1\over\sqrt{2}\sigma_{\cal Q} \,\text{erfc}^{-1}(1/2)}\right)^2\,.
\ee
Therefore, we have for the square root of the largest eigenvalue of $\mathbf \Xi$
\be
\xi_N\approx{1\over\sqrt{2} \,\text{erfc}^{-1}(1/2)}{ f_{N}\over \sigma_{\cal Q}}\,.
\ee
Using Eq.~(\ref{kinrange}) we find the diameter
\be\label{eqn:heavydiam}
{\cal D}\approx {f_{N}\over\sqrt{2} \,\text{erfc}^{-1}(1/2) \sigma_{\cal Q}}  \lVert \resc{\Psi}^{Q^\top Q}_N\rVert\lesssim \begin{cases}f_{N} N~~~~~\,\text{for~}~~~P=N\\ f_{N} \sqrt{N}~~~\text{for~}~~~P>N\end{cases}  \,.
\ee
\begin{figure}
  \centering
  \includegraphics[width=1\textwidth]{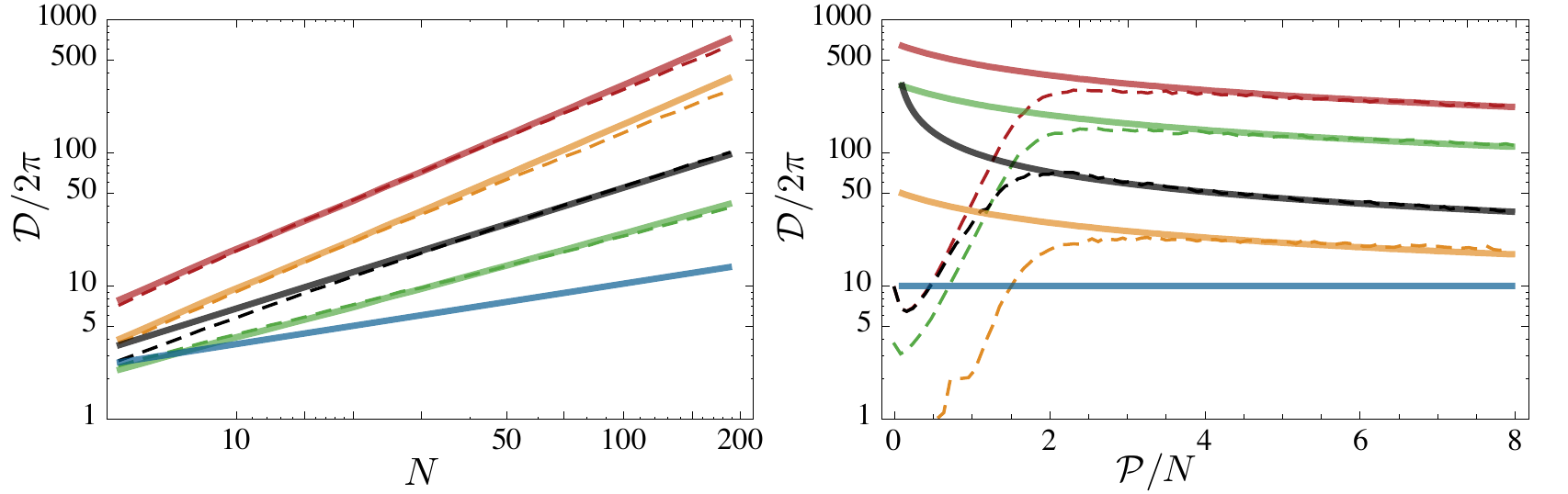}
  \caption{\small Left: Diameter versus the number of fields for a fixed number ${\mathcal P}=4N$ of non-vanishing entries in ${\mathbf Q}$. Right: Kinematic range vs.~${\mathcal P}/N$ for fixed $N=100$. Dashed lines illustrate numeric results, and the solid lines are the analytic results. From top to bottom, red: unit metric~(\ref{eqn:unitdiam}); green: Wishart metric~(\ref{wisrange}); gray: non-square $\mathcal Q$ matrix ~(\ref{wisrange}) with $P-N=3$; orange: heavy-tailed metric~(\ref{eqn:heavydiam});  blue: $\sqrt{N}$ for comparison.\label{fig:transitionplot}}
  \end{figure}
Finally, Figure \ref{fig:transitionplot} illustrates numerically the approach to universality and the scaling of the kinematic range with $N$.

\section{Dynamic Alignment}\label{sec:infdynamics}

So far we have evaluated the typical diameter of the fundamental region, which we found to be parametrically larger than the typical scale of the metric eigenvalues. However, in order to realize large field chaotic inflation within one fundamental domain,\footnote{Inflation could proceed beyond one fundamental domain, as we will discuss,  and could span many fundamental domains in the presence of monodromy.} the diameter in the light directions of the potential is required to be large.\footnote{In \cite{Bachlechner:2014hsa} it was shown that for a trivial $\cal Q$ matrix the direction of largest field space diameter is generically misaligned with respect to the lightest canonical field.} In this section we consider the diameter for a displacement of the lightest canonical field.
We will find that universality generically leads to an alignment of the largest direction with the lightest canonical field.

Let us again consider the Lagrangian~(\ref{eqn:lag}) for the fields $\boldsymbol \phi$.  Well inside the fundamental domain, with $-\pi\ll \left( {\mathbf Q} {\mathbf Q}_R^{-1} \boldsymbol \phi\right)_i\ll\pi$, we can expand the potential to quadratic order,
\be\label{eqn:enhance}
{\mathcal L}={1\over 2} \partial \boldsymbol\phi^\top \mathbf \Xi \, \partial \boldsymbol\phi -{1\over 2} \boldsymbol\phi^\top\mathbf M_\phi^2 \, \boldsymbol\phi\,,
\ee
where
\be
\mathbf M_\phi^2=\diag({\Lambda_{1,\dots N}^4})+(\mathbf Q_\text{R})^\top \diag({\Lambda_{N+1,\dots P}^4)}\, \mathbf Q_\text{R}\,,
\ee
is the mass matrix in the $\phi$ basis. The canonically normalized fields $\boldsymbol \Phi$ are given by
\be
\boldsymbol \Phi=\diag(\xi_i)\, \mathbf S^\top_{\Xi} \, \boldsymbol\phi\,,
\ee
and the Lagrangian becomes
\be\label{eqn:lagPhi}
{\mathcal L}={1\over 2} \partial \boldsymbol \Phi^\top \partial\boldsymbol\Phi-{1\over 2}\boldsymbol \Phi^\top \mathbf M_{\Phi}^2 \boldsymbol \Phi\,,
\ee
where
\be
{\mathbf M}^2_{\Phi}=\diag(1/\xi_i) \, \mathbf S^\top_{\Xi} \, \mathbf M_{\phi}^2 \, \mathbf S^{\phantom{T}}_{\Xi} \, \diag(1/\xi_i)\,.
\ee

To obtain a lower bound on the typical arc length traversed during the approach to the vacuum, we consider a scan over random initial conditions, uniformly distributed over the boundary  of validity of the quadratic approximation,\footnote{Note that most of the volume of an $N$-polytope is concentrated at the boundary, so a scan over initial positions  that is uniform throughout the polytope would yield displacements similar to those from a scan over the boundary.} i.e.~we examine an initial point
\be
\resc{\hat{\mathbf v}},
\ee
where $\hat{\mathbf v}$ is a unit vector with uniform probability density on the sphere $S^{N-1}$.  The semidiameter of the fundamental domain in the direction $\hat{\mathbf v}$
is a lower bound for the dynamical field range, and is given by
\be\label{dynrange}
{1\over 2}{\cal D}_{\hat{\mathbf v}}={1\over 2}\lVert\diag{\xi_i} \, \mathbf S_{\Xi}^\top\resc{\hat{\mathbf v}}\rVert\,.
\ee
Because the initial points $\resc{\hat{\mathbf v}}$  are uniformly distributed on $S^{N-1}$, the displacements $\resc{\hat{\mathbf v}}$ will typically have overlaps of $1/\sqrt{N}$ with the direction corresponding to the maximum diameter of the fundamental domain. Thus, the typical displacement from the vacuum in a scan over random initial conditions is given by
\be\label{dynrange2}
{\cal D}_{\hat{\mathbf v}} \approx {1\over \sqrt{N}}{\cal D}\,.
\ee

In the above estimate we considered the {\it typical} field range when scanning over initial conditions uniformly distributed in the fundamental domain. However,  one might also be interested in the maximum field range over which the quadratic approximation is valid, along the direction of the lightest field.
To analyze this, we assume that the hierarchy in the eigenvalues of the kinetic matrix $\mathbf \Xi$ is  much larger than the hierarchy of the entries in the rotated mass matrix $ \mathbf S^\top_{\Xi}\, \mathbf M_{\phi}^2\,  \mathbf S^{\phantom{T}}_{\Xi}$.\footnote{For the case of a Wishart metric $\mathbf K$ we  have verified numerically that the hierarchy of the entries of $\mathbf S^\top_{\Xi} \mathbf M_{\phi}^2\,  \mathbf S^{\phantom{T}}_{\Xi}$ is parametrically smaller than the hierarchy in the matrix $\xi_i\xi_j$, by a factor of order $N^2$ independent of the $\Lambda_i$, leading to dynamic alignment.} The mass matrix for the  canonically normalized fields $\boldsymbol \Phi$ is  then dominated by the $\xi$ contribution:
\be
\mathbf M_{\Phi}^2=\diag(1/\xi_i) \, \mathbf S^\top_{\Xi}\, \mathbf M_{\phi}^2\, \mathbf S^{\phantom{T}}_{\Xi}\,\diag(1/\xi_i)\approx{ \Lambda_{ M_\phi}^4 \over \xi_i\xi_j}\,,
\ee
where $\Lambda_{M_\phi}^4$ is the typical scale of the entries of $\mathbf S^\top_{\Xi} \, \mathbf M_{\phi}^2 \, \mathbf S^{\phantom{T}}_{\Xi}$, so that the lightest direction is given approximately by
\be
\hat{v}^\Phi={v^\Phi\over |v^\Phi|}\sim \left(0,\dots,0,{1}\right)\,,
\ee
which approximately coincides with the direction giving the maximum diameter.
This alignment occurs because in the $\phi$ basis the light direction corresponds to $\Psi^\Xi_N$, the eigenvector corresponding to the largest axion decay constant.
Using (\ref{dynrange}), we find that the diameter in the direction of the lightest field is
\be\label{kintodyn}
{\cal D}_{\text{light}}\approx {\cal D}\,.
\ee

\begin{figure}
  \centering
  \includegraphics[width=.5\textwidth]{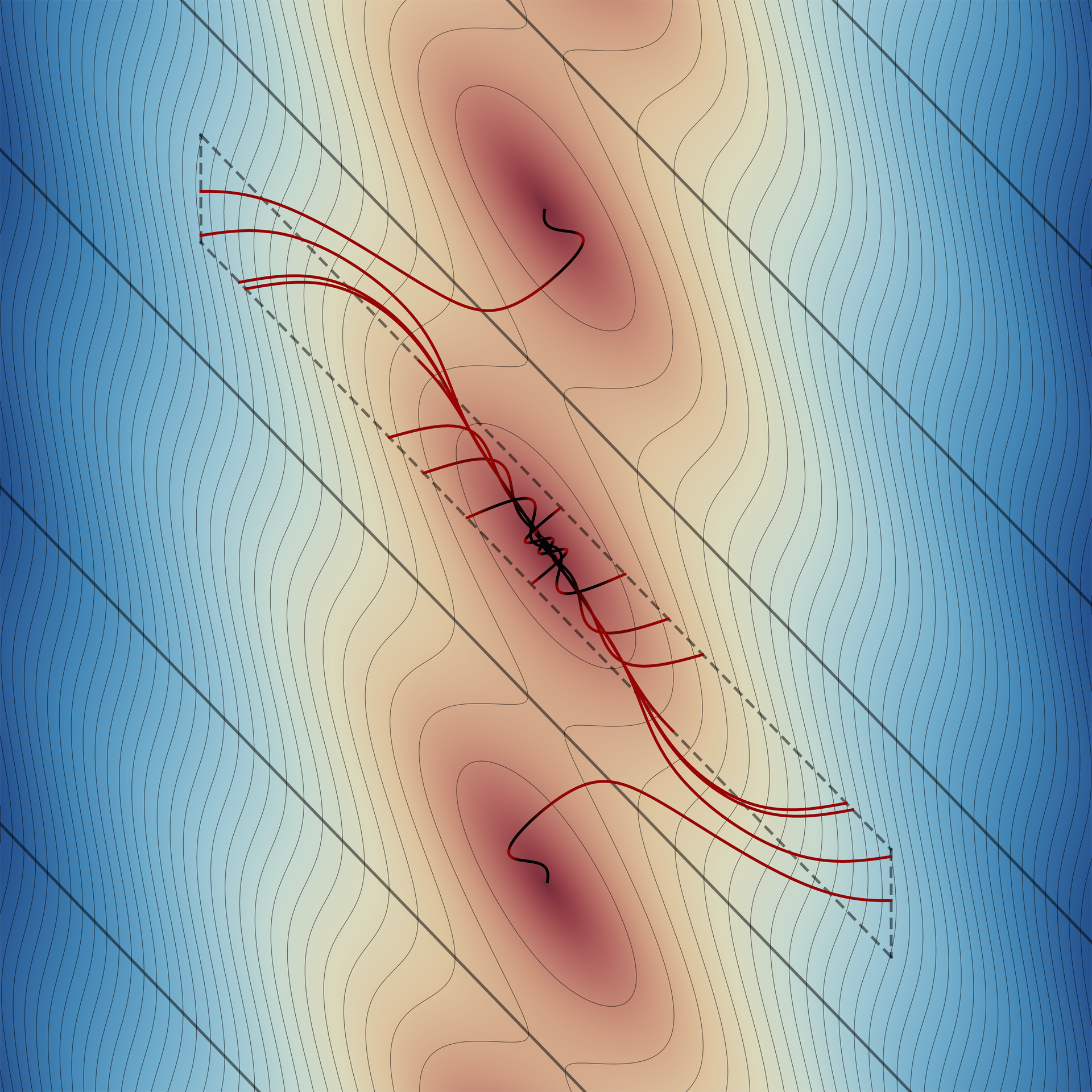} 
  \caption{\small Contour plot of a two-dimensional axion potential, along with the region of validity of the quadratic expansion and a set of randomly chosen inflationary trajectories. The axes are canonically normalized fields.\label{dynamicsfig}}
  \end{figure}

We  have observed a generic enhancement to the diameter of a single fundamental domain of the potential, due to eigenvector delocalization and the nontriviality of the $\mathcal Q$ matrix. This is a  promising setting for realizing chaotic inflation.
Starting the system with a displacement along the lightest direction can lead to single-field slow roll inflation in a quadratic potential:
\be
V(\Phi)={1\over 2} m^2 \Phi^2\, ,
\ee
which yields a large number of $e$-folds,
\be
N_{e}={1\over 4}|\Delta\Phi|^2\gtrsim {1\over 16 N} {\cal D}_{\text{light}}^2\, ,
\ee
where we used the estimate from (\ref{kintodyn}).
For example, taking the metric on moduli space to be a Wishart matrix, we find the scaling
\be
N_{e}\propto N^3 {f_{N}^2\over \M^2}\,.
\ee

Although single-field inflation is a possibility  in this system, it is not a generic outcome.  Instead, the  more massive fields will decay first,  with the lighter fields settling into their minima later.
This process is illustrated in Figure \ref{dynamicsfig}. A number of features are worth noting. While kinetic alignment  allows the diameter of one lattice domain  to be super-Planckian at large $N$,
this does not imply that the inflationary trajectory remains within a region where a quadratic approximation to the potential is valid.
In particular, although the large hierarchy in the axion decay constants leads to an approximate alignment of the least massive direction with the kinematically largest direction, a slight misalignment can lead to an evolution into a neighboring minimum. This does not spoil the possibility of inflation: there is still a large field displacement, and inflation can proceed driven during the approach to the neighboring minimum. These effects, in particular the multifield dynamics during the onset of inflation, can give rise to interesting physical phenomena, such as non-adiabatic perturbations or even domain walls.
A full analysis of these effects is beyond the scope of this work.

\section{Axions in Supergravity}\label{sec:sugra}
Our discussion so far has been at the level of a low-energy effective field theory containing $N$ axions.
However, because high-scale inflation is extremely sensitive to physics at the Planck scale, it is important to inform the effective description with the data of an ultraviolet completion.
We will therefore explain how our considerations extend to axions in string theory.   As a bridge between our general analysis and  specific string theory constructions, we  now discuss axions in  four-dimensional ${\cal N}=1$  supergravity theories, incorporating the structures of the effective  supergravity theories that arise in the flux compactifications of type IIB string theory described in \S\ref{sec:DDFGK}.
The effective supergravities presented here generically exhibit kinetic and dynamic alignment.

\subsection{Hessian matrix}
We will  now examine the scalar potential in an $\mathcal{N} = 1$ supergravity theory, with an eye towards the K\"{a}hler moduli sector of Calabi-Yau compactifications of type IIB string theory. The Lagrangian of the chiral superfields $\phi^A$ is given by
\be
{\mathcal L}=K_{A\bar{B}} (\phi^C,\bar{\phi}^{\bar{D}}) \partial_\mu \phi^A\partial^\mu \bar{\phi}^{\bar{B}}-V(\phi^C,\bar{\phi}^{\bar{D}})\, ,
\ee
with the F-term potential\footnote{We omit the D-term potential, because in the constructions that we will discuss, the D-terms do not involve the axions to leading order, and can be safely ignored in analyses of inflationary dynamics.}
\be
V(\phi^C,\bar{\phi}^{\bar{D}})=e^{K}\left(K^{A\bar{B}} D_A W \bar{D}_{\bar{B}}\overline{W} -3|W|^2\right)\, .
\ee
In the above equations $K_{A\bar{B}}$ is the K\"ahler metric on moduli space,  which is independent of the axions at the perturbative level, and $W$ is the holomorphic superpotential. In the case of type IIB string theory, the indices $A$ and $\bar{B}$ run over the dilaton, the complex structure moduli, and the K\"ahler moduli, such that $A=1, \dots, h^{1,1}+h^{2,1}+1$. As stated before, we will concern ourselves with the case  in which the complex structure moduli and dilaton are integrated out supersymmetrically at a high scale, so we will henceforth restrict ourselves to an effective theory for the K\"ahler moduli $T^j = \tau^j +i\theta^j$, labeled by the indices $i,j$.
A consistency requirement for our analysis is that the motion of the inflaton does not destabilize any fields that we have assumed to be set at their minima. We will therefore examine the cross-coupling terms in the Hessian, and ensure that these are not large enough to push a previously-stable saxion away from its minimum so as to destabilize the configuration.\footnote{A discussion of this problem in the context of N-flation appears in \cite{Kallosh:2007cc}; see also \cite{Grimm:2007hs}.}  At a supersymmetric critical point we can write the potential in terms of small fluctuations as
\be
V(T,\bar{T})=V(T_0)+\sum_{ij}\partial_i\partial_j VT^iT^j=V(T_0)+\left(\begin{array}{c c}\bar T &T\end{array}\right) {\cal H}\left( \begin{array}{c}T \\\bar T\end{array}\right)
\,,
\ee
where $i,j$ run over unbarred and barred indices and $T$ denotes the fluctuations about the minimum. The Hessian matrix is given by
\bea
{\cal H} &=&
\left(
\begin{array}{c c}
\partial^2_{i \bj} V & \partial^2_{i j} V \\
\partial^2_{\bi \bj} V & \partial^2_{\bi j} V
\end{array}
\right)
=
{\cal H}_Z
 - 2 |W|^2 \left(
\begin{array}{c c}
K_{i\bj} &0\\
0 &K_{\bi j}
\end{array}
\right) \,  , \label{eq:Hess}
\eea
where
\bea
{\cal H}_Z =
\left(
\begin{array}{c c}
 Z_{i}^{~\bar A}\ \bZ_{\bi \bar A}  &  - Z_{i j} \bW \\
 - \bZ_{\bi \bj} W & \bZ_{\bi}^{~A}\ Z_{j A}
\end{array}
\right)  \, , \label{Q}
\eea
and $Z_{AB} = Z_{BA}\equiv {\mathcal D}_A D_B W$.  Here ${\mathcal D}_AV_B=\partial_AV_B+K_AV_B-\Gamma_{AB}^CV_C$, and we have used K\"ahler transformations to set $K=0$ at the critical point.

We can transform the Hessian matrix into a $(\tau~\theta)$ basis via
\be\label{basistrafo}
\left( \begin{array}{c}T \\\bar T\end{array}\right)=\left( \begin{array}{c c}\mathbb{1}& \mathbb{1}i\\\mathbb{1}&-\mathbb{1}i\end{array}\right)\left( \begin{array}{c}\tau \\\theta\end{array}\right)= \mathbf U\left( \begin{array}{c}\tau \\\theta\end{array}\right)\,,
\ee
such that
\be
V(\tau,\theta)=V(\tau_0)+\left(\begin{array}{c c}\tau &\theta \end{array}\right) \mathbf U^\dagger{\cal H} \mathbf U \left( \begin{array}{c}\tau \\\theta\end{array}\right)\,.
\ee
We then have
\be\label{eqn:axionsaxionhess2}
{\mathcal H}^{\tau\theta} = \mathbf U^\dagger{\cal H}\mathbf U \, ,
\ee
which evaluates to
\be\label{eqn:axionsaxionhess}
{\mathcal H}^{\tau\theta}= 2\left( \begin{array}{c c}Z\bar{Z}-2|W|^2  K -{1\over 2}\left(\overline{W}Z+W \bar{Z}\right)&{i\over 2}\left(\overline{W} Z-W \bar{Z}\right)\\{i\over 2}\left(\overline{W}Z-W \bar{Z}\right)&\bar{Z} Z-2|W|^2 f K +{1\over 2}\left(\overline{W} Z+W \bar{Z}\right)\end{array}\right) \,.
\ee
Here $Z \bar{Z}$ is contracted using the K\"ahler metric. Let us now consider the couplings between the saxions $\tau^i$ and the axions $\theta^i$. In \cite{Bachlechner:2012at} it was shown that  tachyons allowed by the Breitenlohner-Freedman bound are ubiquitous in AdS vacua, and will render an uplifted solution unstable, unless $|W|\ll m_{susy}/N$. Here, $m_{susy}$ is the scale of the supersymmetric fermion mass matrix $Z_{ij}$. Therefore, the scale of the masses of $\tau^i$ is given by $ Z\bar{Z}\sim M_{\tau^2}^2\sim m_{susy}^2$, while the couplings between $\tau$ and $\theta$ are given by $M^2_{\tau\theta}\sim W\bar{Z}\sim W m_{susy}$. Then the constraint $|W|\ll m_{susy}/N$ leads to
\be
M_{\tau^2}^2\gg{1\over N} M^2_{\tau\theta}\,, .
\ee
To leading order in $\tau$ and $\theta$, the displacement of the minimum for $\tau$ can be estimated by solving $\partial_{\tau} V(\tau,\theta)|_{\tau=\tau_{min}}=0$, which gives
\be
\lVert \Delta\tau_{min}\rVert=\lVert \left(M^2_{\tau\tau}\right)^{-1}M_{\tau \theta } \Delta\theta \rVert\sim {1\over N} \lVert \Delta \theta\rVert\,.
\ee
Here we have considered only the leading order contributions to the $\tau$-$\theta$ mixing terms in the Hessian. In general there will be higher-order contributions, but when our expansion is valid these are not large enough to destabilize the vacuum.

We now turn to a more specific effective supergravity theory, in which the superpotential takes the form
\be\label{eqn:gensup}
W= W(S,\chi)+\sum_j A_j(\chi_a) e^{-q^j_{~i}T^i}\, = W(S,\chi)+\sum_j A_j(\chi_a) e^{-q^j_{~i}(\tau^i+i\theta^i)}\,.
\ee
In the last equality we have expressed the complex chiral scalar in terms of its real saxion and axion components. If the K\"ahler potential is independent of the axions, at least to the order at which we are working,  then the axion potential can be written
\be\label{effectivepot}
V = C + \sum\limits_{j} B_j\, \text{cos}(q^j_{~i}\,\theta^i - \theta_W)
 + \sum\limits_{j <k} B_{jk}\, \text{cos}(q^j_{~i}\theta^i - q^k_{~i}\theta^i)\, .
\ee
In this formula, $C, B_j$, and $B_{jk}$ depend on the saxions but not on the axions. In \S\ref{sec:DDFGK}, we will consider the KKLT moduli stabilization scheme in type IIB string theory, which requires solving the F-flatness constraints $F_i = 0$, $\forall \, i$. In general the $A_i$ prefactors in each nonperturbative term will be complex, and will contribute a phase to each exponential. When we have $N$ axions we can simply perform a shift to absorb each $A_i$ phase, and can therefore take the $A_i$ to be real. In addition, we can perform a K\"{a}hler transformation to make $W_0$ real and negative. For the remainder of this work we will assume that these transformations have been performed.

To extract the axion-saxion coupling at the supersymmetric minimum we need to compute the matrix $Z_{AB} = \mathcal{D}_A D_B W$, where $\mathcal{D}_A$ is the geometrically covariant and K\"ahler covariant derivative, and $D_B$ is the K\"ahler covariant derivative. At a supersymmetric minimum $D_A W \equiv F_A = 0$, so we can write
\be
Z_{AB} = \mathcal{D}_AD_B W = \partial_A F_B+ K_{,A}F_B + \Gamma^{C}_{AB}F_C = \partial_A F_B,\ .
\ee
Writing $F_B = (\partial_B + K_{,B})W$, we have
\be
Z_{AB} = \partial^2_{AB}W + K_{,B}\partial_A W + K_{,AB}W\, .
\ee
This is not manifestly symmetric in the induced $A$ and $B$, but we can fix that by multiplying the critical point equation by $K_{,A}$:
\be
K_{,A}\partial_BW = -K_{,A}K_{,B}W\, .
\ee
Therefore, we find
\be
Z_{AB} = \partial^2_{AB}W +K_{,AB}W -K_{,A}K_{,B}W\, .
\ee
Applying this to (\ref{eqn:gensup}) we find
\be
Z_{ij} = \sum\limits_{k}A_k \left(q^k_{~i}\ q^k_{~j} \right) e^{-q^k_{~i}T^{i}} + \left(K_{,ij} -K_{,i}K_{,j}\right)W\, .
\ee
The scale of the inflaton mass is approximately set by the scale $m_{susy}/N$. If the axions are stabilized at $\theta^i = 0$, then the superpotential will be real at the minimum, as will the matrix $\mathbf Z$. Therefore, from the form of equation (\ref{eqn:axionsaxionhess}), the axions and the saxions will be decoupled to leading order, and we do not need to worry about destabilizing the saxions during inflation, as long as each axion does not move too much. For this reason we will focus on the $\theta^i = 0$ vacuum.

\subsection{Approach to universality}

\begin{figure}
  \centering
  \includegraphics[width=.5\textwidth]{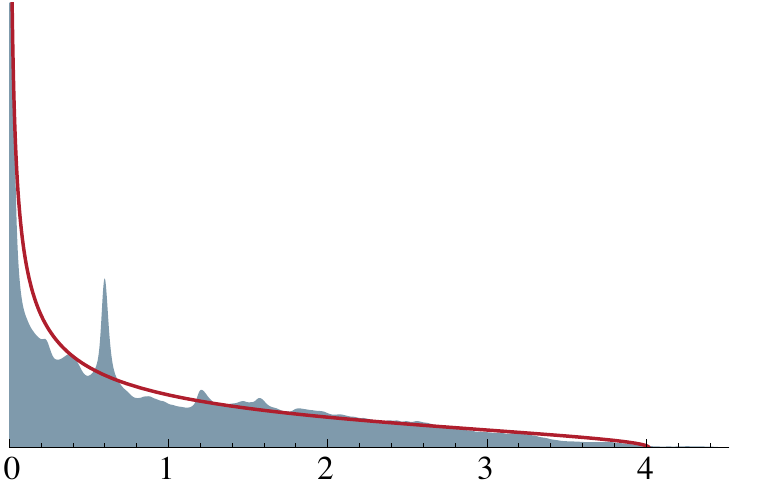}
  \caption{\small Normalized probability distribution of the eigenvalue spectrum of $\mathbf Q^\top \mathbf Q$ along with the analytic Wishart eigenvalue spectrum.
\label{figQ}}
\end{figure}

The full effective potential in (\ref{effectivepot}) has $P$ cosine terms appearing due to the nonperturbative superpotential and an additional $N^2$ terms of the form $\text{cos}(\phi^i - \phi^j)$, appearing as cross terms with $\mathbf Q$ matrix $\mathbf Q_{\text{cross}}$. The full ${\cal Q}$ matrix is then given as
\be
{\cal Q}=\left( \begin{array}{c} \mathbf Q\\ \mathbf Q_{\text{cross}}\end{array}\right)\,.
\ee
Note that the additional constraints on the fundamental domain originating from the cross terms decrease the diameter found by considering only superpotential periodicities by at most a factor of $2$, because only differences $\phi^i - \phi^j$ appear. Therefore, the cross terms contain no new physical enhancement of, or limitation on, the diameter of the fundamental domain. However, the effective potential contains the full matrix ${\cal Q}$ and picking an arbitrary full rank $N\times N$ matrix can be used to define the axions. The metric on field space and its decay constants, however, do depend on the choice of axions. In particular, because there is a large number of possible full rank matrices, with essentially random entries, the metric on field space approaches that of an inverse Wishart matrix, independent of the periodicities in the nonperturbative superpotential. This approach to universality is illustrated in Figure \ref{figQ}. Here we chose $\mathbf Q={\mathbb 1}$, $K={\mathbb 1}f^2$, $N=51$ and defined the axions $\boldsymbol \phi=\mathbf Q \, \boldsymbol \theta$ in terms of a full rank matrix $\mathbf Q$ that consists of $N$ randomly chosen rows of the full matrix ${\cal Q}$. Due to universality, the metric on moduli space approaches an inverse Wishart distribution with potentially large eigenvalues. This observation is purely due to the fact that the definition of the axions and the associated metric is arbitrary. Despite the presence of very large metric eigenvalues, in this example the field range is not enhanced compared to the trivial case ${\cal Q}={\mathbb 1}$. This is a consequence of the fact that the axion lattice domains are defined by the periodicities of the superpotential.

\section{Diameter in an Explicit String Compactification}\label{sec:DDFGK}

It will be instructive to verify that the kinetic alignment mechanism we have described can occur in a UV-complete theory, at large $N$.
Weakly-coupled string theory is, at the moment, our best tool for testing whether a particular mechanism is consistent with a theory of quantum gravity. 
In this section we will discuss an explicit compactification of type IIB string theory with moderately large $N$ and a nontrivial ${\mathbf Q}$ matrix. Our findings suggest that the kinetic alignment discussed above can occur very naturally in compactifications of type IIB string theory on certain Calabi-Yau orientifolds.

We will examine a state-of-the-art string compactification, with $h^{1,1}=51$, that was introduced by Denef, Douglas, Florea, Grassi, and Kachru (DDFGK)~\cite{Denef:2005mm}.  Their construction is almost completely explicit: quantized flux values are specified to stabilize the complex structure moduli and dilaton at weak coupling, and the K\"{a}hler moduli are stabilized by nonperturbative effects, which are known to be present and to provide non-vanishing contributions to the superpotential. The only piece that is not completely explicit are the Pfaffian prefactors of the nonperturbative superpotential terms, which are set to unity.

\subsection{Axions in type IIB string theory}
Type IIB string theory compactified on a Calabi-Yau threefold $X_3$ yields an $\mathcal{N}=2$ $d=4$ effective theory. In the absence of branes, the massless fields are the $h^{2,1}$ vector multiplets, which include the complex structure moduli, and $h^{1,1}$  hypermultiplets, which include the K\"{a}hler moduli. We are interested in a $\mathcal{N}=1$ theory, which can be obtained by orientifolding, resulting in an $\mathcal{N}=1$ supergravity theory with an internal space $\hat{X}_3$, the orientifold of the threefold $X_3$.  
For simplicity we will assume that all of the divisors are even under the orientifold action (general at present, but specified in the example of \S\ref{sec:ddfgk}). 
The complex structure moduli are lifted by a tree-level Gukov-Vafa-Witten flux superpotential~\cite{Gukov:1999ya}, while the K\"{a}hler moduli are massless at leading order, due to the shift symmetry of the imaginary part of the K\"{a}hler moduli.  The perturbative continuous shift symmetries are broken to discrete shifts by nonperturbative effects, such as Euclidean D3-branes wrapping internal four-cycles, or gaugino condensation on stacks of D7-branes wrapping such cycles. At large volume, the masses of the complex structure moduli are hierarchically larger than those of the K\"{a}hler moduli, so that the complex structure moduli can typically be integrated out, yielding an effective theory for the K\"{a}hler moduli.

At large volume, the leading order action is determined by the classical K\"{a}hler potential for the K\"ahler moduli, and by the leading order contributions to the nonperturbative superpotential. The classical K\"{a}hler potential takes the form
\be
K = -2 \ \text{log}(V), \quad V = \frac{1}{6}\int J \wedge J \wedge J\,.
\ee
Here the K\"{a}hler form is expanded as $J = t^i \omega_i$, where $\omega_i \in H^{1,1}(X, \mathbb{Z})$. The K\"{a}hler moduli have a natural interpretation as the volumes of four-cycles. These volumes combine with periods of the Ramond-Ramond four-form to form chiral superfields. The complex scalar components take the form:
\be
T^j = \frac{1}{2}\int\limits_{D_j}J \wedge J + i\int\limits_{D_j} C_{(4)} \equiv \tau^j + i \theta^j\,.
\ee
We will write the general nonperturbative superpotential as
\be
W = W_0 + \sum\limits_i A_i e^{- q^{i}_{\ j}T^j}\,.
\ee
Here $W_0$ is the value of the flux superpotential with the complex structure fields set at their minima, and the $A_i$ are one-loop determinants.

\subsection{The compactification}\label{sec:ddfgk}
The geometry (before orientifolding) is a resolution of the orbifold $T^6/\mathbb{Z}_2 \times \mathbb{Z}_2$, which has 51 K\"ahler moduli and 3 complex structure moduli. $T^6 = (T^2)^3$ has three K\"{a}hler moduli, which descend to the so-called ``sliding divisors" $\{R_i\}, i = 1\dots 3$. The orbifold action is
\begin{center}
$\begin{matrix}
 & & z_1 & z_2 & z_3 \\
\alpha&  & + & - & - \\
\beta &  & - & + & - \\
\alpha \circ \beta &  & - & - & + \\
\end{matrix}$
\end{center}

There are 48 fixed lines under the orbifold action, whose resolution introduces 48 exceptional divisors, denoted by $\{E_{i\alpha,j\beta}\}$, where $i = 1 \dots 3, \alpha = 1\dots 4, i < j$. We will consider what DDFGK refer to as the ``symmetric resolution." There are 12 fixed divisors under the orientifold action, resulting in 12 O7-planes. An SO(8) stack of D7-branes is placed on each O7-plane. The D7-brane divisors will be denoted by $D_{i\alpha}$. In the compact model the $D_{i\alpha}$ are disjoint, so there is no massless bifundamental matter arising from intersections of D7-branes. In addition, the $D_{i\alpha}$ are rigid, so there is no adjoint matter, and the gauginos will condense. The $D_{i\alpha}$ can be expressed in terms of the sliding divisors and exceptional divisors. For example,
\begin{equation}
D_{1\alpha} = R_1 - \sum\limits_{\beta}E_{1\alpha,2\beta} -  \sum\limits_{\gamma}E_{3\gamma,1\alpha}\,.
\end{equation}
Each exceptional divisor is rigid, and supports a Euclidean D3-brane, which generates a superpotential of the form
\begin{equation}
\Delta W \sim e^{-2\pi\tau_{i\alpha,j\beta}}\,.
\end{equation}
The gaugino condensates generate superpotentials of the form
\begin{equation}
\Delta W \sim e^{-2\pi\tau_{i\alpha}/6}\,,
\end{equation}
where we have used the fact that the dual Coxeter number of SO(8) is 6. Expanding the K\"ahler form as
\begin{equation}
J = r_i R_i - t_{1\alpha,2\beta}E_{1\alpha,2\beta}- t_{2\beta,3\gamma}E_{2\beta,3\gamma}- t_{3\gamma,1\alpha}E_{3\gamma,1\alpha}\,,
\end{equation}
the volume of the orientifold can be written as
\begin{eqnarray}
V = r_1 r_2 r_3 -\frac{1}{2}\left(r_i\sum\limits_{\beta \gamma} t^2_{2\beta,3\gamma} + \dots \right) -\frac{1}{3}\left(\sum\limits_{\alpha\beta}t^3_{1\alpha,2\beta}+\dots \right) \nonumber \\
+\frac{1}{4}\left(\sum\limits_{\alpha \beta \gamma} t_{1\alpha,2\beta}t^2_{2\beta,3\gamma} + t_{1\alpha,2\beta}t^2_{3\gamma,1\alpha} + \dots\right) -\frac{1}{2}\sum\limits_{\alpha\beta\gamma}t_{1\alpha,2\beta}t_{2\beta,3\gamma}t_{3\gamma,1\alpha}\,.
\end{eqnarray}
The areas of the generators of the Mori cone are
\begin{eqnarray}
A_{i,j\beta} = r_i - \sum\limits_{\alpha}t_{i\alpha,j\beta}\, , \nonumber \\
A^{++-} = \frac{1}{2}\left(t_{1\alpha,2\beta} + t_{2\beta,3\gamma} - t_{3\gamma,1\alpha}\right)\,,
\end{eqnarray}
plus cyclic permutations of the latter. DDFGK found a particularly symmetric critical point by setting
\begin{equation}
t_{i\alpha,j\beta} = t, \quad r_i = r\,,
\end{equation} through which the curve areas and divisor volumes simplify greatly:
\begin{eqnarray}
V = r^3 -24rt^2 + 48t^3\, , \nonumber \\
V_{i\alpha,j\beta} = V_E = rt - 3t^2\, , \nonumber \\
V_{i\alpha} = V_D = r^2 -8rt + 16t^2\, , \nonumber \\
A_{i,j\beta} = A_r = r-4t\, , \nonumber \\
A_{\alpha\beta\gamma} = A_t  = \frac{t}{2}\,.
\end{eqnarray}
Under the assumption that the one-loop determinants can be set to unity, a minimum was sought where the phases vanish. The superpotential can then be written as
\begin{equation}
W = W_0 +48e^{-2\pi\left(tr - 3t^2\right)} +12 e^{-2\pi\left(r^2-8rt+16t^2\right)/6}\,.
\end{equation}
DDFGK explicitly stabilized the complex structure moduli using flux, finding that $W_0 \sim -0.3$, which gives a supersymmetric local minimum with the K\"ahler parameters
\begin{equation}
r \approx 4, \quad t \approx 0.4,
\end{equation}
yielding volumes of
\begin{equation}
V \approx 55, \quad V_E \approx 1, \quad V_D \approx 6, \quad A_r \approx 2.5, \quad A_t \approx 0.2\,.
\end{equation}
These values are not parametrically large, and one should ask whether additional perturbative and nonperturbative effects are important in this regime of parameters.
DDFGK directly demonstrated that the leading  known corrections are controllably small, as we now explain.
There are nonperturbative corrections to both the K\"{a}hler potential and the superpotential.
There could be a contribution to the superpotential from multi-wrapped or fluxed instantons, but these contributions will be suppressed by higher-order powers of the exponential  that is already present. Since the values of these exponentials are $e^{-2\pi V_E} \sim 5 \times 10^{-4}$ and $e^{-2\pi V_D/6} \sim 2\times 10^{-3}$, these contributions are expected to  shift the minimum by a very small amount. The corrections to the K\"ahler potential are a bit more complicated, especially given the small volumes of the exceptional curves. First, there are perturbative $\alpha'$ effects, which correct the K\"ahler potential to
\be\label{kahlerpot}
K = -2 \, \text{log}\left(V + \frac{\xi}{g_s^{3/2}} \right), \quad \xi \equiv -\frac{\chi (Y) \zeta (3)}{8 (2\pi)^3} \approx -0.06\,.
\ee
In this formula, $\chi (Y)$ is the Euler characteristic of the ``upstairs" Calabi-Yau. This correction gives  a percent-level correction to the volume, and can therefore be consistently neglected. Nonperturbative corrections can be estimated through the corresponding correction to the underlying $\mathcal{N} =2$ prepotential:
\be
\Delta \mathcal{F} \approx \frac{1}{(2\pi)^3} \sum\limits_{n=1}^{\infty} \frac{1}{n^3}e^{-2\pi\sqrt{g_s}(2 A_t)n} \approx 10^{-3}\,.
\ee
Here we have restricted to a sum over worldsheets wrapping exceptional curves, since these will give the leading order contribution. We have also included a factor of two relevant in moving from the upstairs space to the downstairs space. There are 192 minimal exceptional curves, which in turn provide a percent-level correction to the K\"ahler potential. In addition, since $g_s\approx 0.27$ is moderately small, string loop corrections should not significantly shift the minimum.  More details on these results  can be found in~\cite{Denef:2005mm}.

We will consider this point in moduli space as a toy model. It is, of course, not a realistic model for inflation, as the minimum is a supersymmetric AdS vacuum.  However, it is still instructive to demonstrate kinetic alignment in a completely explicit  and well-controlled string compactification.
To compare this example to the rest of the paper, we write the nonperturbative contributions  to the superpotential in the form $\sum_i A_i e^{- q^{i}_{\ j}T^j}$, so that the eigenvalues of the kinetic matrix correspond to axion decay constants. The matrix $\mathbf q$ is then given by

\begin{flalign}\label{fullq}
{\mathbf q}&=
\setstretch{.5}
\arraycolsep=.8pt
{\fontsize{.3cm}{1.2em}\selectfont {\pi\over 3}
\left(
\begin{array}{ccccccccccccccccccccccccccccccccccccccccccccccccccc}
 \one & 0 & 0 &  \sm\one &  \sm\one &  \sm\one &  \sm\one & 0 & 0 & 0 & 0 & 0 & 0 & 0 & 0 & 0 & 0 & 0 & 0 & 0 & 0 & 0 & 0 & 0 & 0 & 0 & 0 & 0 & 0 & 0 & 0 & 0 & 0 & 0 & 0 &  \sm\one &  \sm\one &  \sm\one &  \sm\one & 0 & 0 & 0 & 0 & 0 & 0 & 0 & 0 & 0 & 0 & 0 & 0 \\
 \one & 0 & 0 & 0 & 0 & 0 & 0 &  \sm\one &  \sm\one &  \sm\one &  \sm\one & 0 & 0 & 0 & 0 & 0 & 0 & 0 & 0 & 0 & 0 & 0 & 0 & 0 & 0 & 0 & 0 & 0 & 0 & 0 & 0 & 0 & 0 & 0 & 0 & 0 & 0 & 0 & 0 &  \sm\one &  \sm\one &  \sm\one &  \sm\one & 0 & 0 & 0 & 0 & 0 & 0 & 0 & 0 \\
 \one & 0 & 0 & 0 & 0 & 0 & 0 & 0 & 0 & 0 & 0 &  \sm\one &  \sm\one &  \sm\one &  \sm\one & 0 & 0 & 0 & 0 & 0 & 0 & 0 & 0 & 0 & 0 & 0 & 0 & 0 & 0 & 0 & 0 & 0 & 0 & 0 & 0 & 0 & 0 & 0 & 0 & 0 & 0 & 0 & 0 &  \sm\one &  \sm\one &  \sm\one &  \sm\one & 0 & 0 & 0 & 0 \\
 \one & 0 & 0 & 0 & 0 & 0 & 0 & 0 & 0 & 0 & 0 & 0 & 0 & 0 & 0 &  \sm\one &  \sm\one &  \sm\one &  \sm\one & 0 & 0 & 0 & 0 & 0 & 0 & 0 & 0 & 0 & 0 & 0 & 0 & 0 & 0 & 0 & 0 & 0 & 0 & 0 & 0 & 0 & 0 & 0 & 0 & 0 & 0 & 0 & 0 &  \sm\one &  \sm\one &  \sm\one &  \sm\one \\
 0 & \one & 0 &  \sm\one &  \sm\one &  \sm\one &  \sm\one & 0 & 0 & 0 & 0 & 0 & 0 & 0 & 0 & 0 & 0 & 0 & 0 &  \sm\one &  \sm\one &  \sm\one &  \sm\one & 0 & 0 & 0 & 0 & 0 & 0 & 0 & 0 & 0 & 0 & 0 & 0 & 0 & 0 & 0 & 0 & 0 & 0 & 0 & 0 & 0 & 0 & 0 & 0 & 0 & 0 & 0 & 0 \\
 0 & \one & 0 & 0 & 0 & 0 & 0 &  \sm\one &  \sm\one &  \sm\one &  \sm\one & 0 & 0 & 0 & 0 & 0 & 0 & 0 & 0 & 0 & 0 & 0 & 0 &  \sm\one &  \sm\one &  \sm\one &  \sm\one & 0 & 0 & 0 & 0 & 0 & 0 & 0 & 0 & 0 & 0 & 0 & 0 & 0 & 0 & 0 & 0 & 0 & 0 & 0 & 0 & 0 & 0 & 0 & 0 \\
 0 & \one & 0 & 0 & 0 & 0 & 0 & 0 & 0 & 0 & 0 &  \sm\one &  \sm\one &  \sm\one &  \sm\one & 0 & 0 & 0 & 0 & 0 & 0 & 0 & 0 & 0 & 0 & 0 & 0 &  \sm\one &  \sm\one &  \sm\one &  \sm\one & 0 & 0 & 0 & 0 & 0 & 0 & 0 & 0 & 0 & 0 & 0 & 0 & 0 & 0 & 0 & 0 & 0 & 0 & 0 & 0 \\
 0 & \one & 0 & 0 & 0 & 0 & 0 & 0 & 0 & 0 & 0 & 0 & 0 & 0 & 0 &  \sm\one &  \sm\one &  \sm\one &  \sm\one & 0 & 0 & 0 & 0 & 0 & 0 & 0 & 0 & 0 & 0 & 0 & 0 &  \sm\one &  \sm\one &  \sm\one &  \sm\one & 0 & 0 & 0 & 0 & 0 & 0 & 0 & 0 & 0 & 0 & 0 & 0 & 0 & 0 & 0 & 0 \\
 0 & 0 & \one & 0 & 0 & 0 & 0 & 0 & 0 & 0 & 0 & 0 & 0 & 0 & 0 & 0 & 0 & 0 & 0 &  \sm\one &  \sm\one &  \sm\one &  \sm\one & 0 & 0 & 0 & 0 & 0 & 0 & 0 & 0 & 0 & 0 & 0 & 0 &  \sm\one &  \sm\one &  \sm\one &  \sm\one & 0 & 0 & 0 & 0 & 0 & 0 & 0 & 0 & 0 & 0 & 0 & 0 \\
 0 & 0 & \one & 0 & 0 & 0 & 0 & 0 & 0 & 0 & 0 & 0 & 0 & 0 & 0 & 0 & 0 & 0 & 0 & 0 & 0 & 0 & 0 &  \sm\one &  \sm\one &  \sm\one &  \sm\one & 0 & 0 & 0 & 0 & 0 & 0 & 0 & 0 & 0 & 0 & 0 & 0 &  \sm\one &  \sm\one &  \sm\one &  \sm\one & 0 & 0 & 0 & 0 & 0 & 0 & 0 & 0 \\
 0 & 0 & \one & 0 & 0 & 0 & 0 & 0 & 0 & 0 & 0 & 0 & 0 & 0 & 0 & 0 & 0 & 0 & 0 & 0 & 0 & 0 & 0 & 0 & 0 & 0 & 0 &  \sm\one &  \sm\one &  \sm\one &  \sm\one & 0 & 0 & 0 & 0 & 0 & 0 & 0 & 0 & 0 & 0 & 0 & 0 &  \sm\one &  \sm\one &  \sm\one &  \sm\one & 0 & 0 & 0 & 0 \\
 0 & 0 & \one & 0 & 0 & 0 & 0 & 0 & 0 & 0 & 0 & 0 & 0 & 0 & 0 & 0 & 0 & 0 & 0 & 0 & 0 & 0 & 0 & 0 & 0 & 0 & 0 & 0 & 0 & 0 & 0 &  \sm\one &  \sm\one &  \sm\one &  \sm\one & 0 & 0 & 0 & 0 & 0 & 0 & 0 & 0 & 0 & 0 & 0 & 0 &  \sm\one &  \sm\one &  \sm\one &  \sm\one \\
 0 & 0 & 0 & \six & 0 & 0 & 0 & 0 & 0 & 0 & 0 & 0 & 0 & 0 & 0 & 0 & 0 & 0 & 0 & 0 & 0 & 0 & 0 & 0 & 0 & 0 & 0 & 0 & 0 & 0 & 0 & 0 & 0 & 0 & 0 & 0 & 0 & 0 & 0 & 0 & 0 & 0 & 0 & 0 & 0 & 0 & 0 & 0 & 0 & 0 & 0 \\
 0 & 0 & 0 & 0 & \six & 0 & 0 & 0 & 0 & 0 & 0 & 0 & 0 & 0 & 0 & 0 & 0 & 0 & 0 & 0 & 0 & 0 & 0 & 0 & 0 & 0 & 0 & 0 & 0 & 0 & 0 & 0 & 0 & 0 & 0 & 0 & 0 & 0 & 0 & 0 & 0 & 0 & 0 & 0 & 0 & 0 & 0 & 0 & 0 & 0 & 0 \\
 0 & 0 & 0 & 0 & 0 & \six & 0 & 0 & 0 & 0 & 0 & 0 & 0 & 0 & 0 & 0 & 0 & 0 & 0 & 0 & 0 & 0 & 0 & 0 & 0 & 0 & 0 & 0 & 0 & 0 & 0 & 0 & 0 & 0 & 0 & 0 & 0 & 0 & 0 & 0 & 0 & 0 & 0 & 0 & 0 & 0 & 0 & 0 & 0 & 0 & 0 \\
 0 & 0 & 0 & 0 & 0 & 0 & \six & 0 & 0 & 0 & 0 & 0 & 0 & 0 & 0 & 0 & 0 & 0 & 0 & 0 & 0 & 0 & 0 & 0 & 0 & 0 & 0 & 0 & 0 & 0 & 0 & 0 & 0 & 0 & 0 & 0 & 0 & 0 & 0 & 0 & 0 & 0 & 0 & 0 & 0 & 0 & 0 & 0 & 0 & 0 & 0 \\
 0 & 0 & 0 & 0 & 0 & 0 & 0 & \six & 0 & 0 & 0 & 0 & 0 & 0 & 0 & 0 & 0 & 0 & 0 & 0 & 0 & 0 & 0 & 0 & 0 & 0 & 0 & 0 & 0 & 0 & 0 & 0 & 0 & 0 & 0 & 0 & 0 & 0 & 0 & 0 & 0 & 0 & 0 & 0 & 0 & 0 & 0 & 0 & 0 & 0 & 0 \\
 0 & 0 & 0 & 0 & 0 & 0 & 0 & 0 & \six & 0 & 0 & 0 & 0 & 0 & 0 & 0 & 0 & 0 & 0 & 0 & 0 & 0 & 0 & 0 & 0 & 0 & 0 & 0 & 0 & 0 & 0 & 0 & 0 & 0 & 0 & 0 & 0 & 0 & 0 & 0 & 0 & 0 & 0 & 0 & 0 & 0 & 0 & 0 & 0 & 0 & 0 \\
 0 & 0 & 0 & 0 & 0 & 0 & 0 & 0 & 0 & \six & 0 & 0 & 0 & 0 & 0 & 0 & 0 & 0 & 0 & 0 & 0 & 0 & 0 & 0 & 0 & 0 & 0 & 0 & 0 & 0 & 0 & 0 & 0 & 0 & 0 & 0 & 0 & 0 & 0 & 0 & 0 & 0 & 0 & 0 & 0 & 0 & 0 & 0 & 0 & 0 & 0 \\
 0 & 0 & 0 & 0 & 0 & 0 & 0 & 0 & 0 & 0 & \six & 0 & 0 & 0 & 0 & 0 & 0 & 0 & 0 & 0 & 0 & 0 & 0 & 0 & 0 & 0 & 0 & 0 & 0 & 0 & 0 & 0 & 0 & 0 & 0 & 0 & 0 & 0 & 0 & 0 & 0 & 0 & 0 & 0 & 0 & 0 & 0 & 0 & 0 & 0 & 0 \\
 0 & 0 & 0 & 0 & 0 & 0 & 0 & 0 & 0 & 0 & 0 & \six & 0 & 0 & 0 & 0 & 0 & 0 & 0 & 0 & 0 & 0 & 0 & 0 & 0 & 0 & 0 & 0 & 0 & 0 & 0 & 0 & 0 & 0 & 0 & 0 & 0 & 0 & 0 & 0 & 0 & 0 & 0 & 0 & 0 & 0 & 0 & 0 & 0 & 0 & 0 \\
 0 & 0 & 0 & 0 & 0 & 0 & 0 & 0 & 0 & 0 & 0 & 0 & \six & 0 & 0 & 0 & 0 & 0 & 0 & 0 & 0 & 0 & 0 & 0 & 0 & 0 & 0 & 0 & 0 & 0 & 0 & 0 & 0 & 0 & 0 & 0 & 0 & 0 & 0 & 0 & 0 & 0 & 0 & 0 & 0 & 0 & 0 & 0 & 0 & 0 & 0 \\
 0 & 0 & 0 & 0 & 0 & 0 & 0 & 0 & 0 & 0 & 0 & 0 & 0 & \six & 0 & 0 & 0 & 0 & 0 & 0 & 0 & 0 & 0 & 0 & 0 & 0 & 0 & 0 & 0 & 0 & 0 & 0 & 0 & 0 & 0 & 0 & 0 & 0 & 0 & 0 & 0 & 0 & 0 & 0 & 0 & 0 & 0 & 0 & 0 & 0 & 0 \\
 0 & 0 & 0 & 0 & 0 & 0 & 0 & 0 & 0 & 0 & 0 & 0 & 0 & 0 & \six & 0 & 0 & 0 & 0 & 0 & 0 & 0 & 0 & 0 & 0 & 0 & 0 & 0 & 0 & 0 & 0 & 0 & 0 & 0 & 0 & 0 & 0 & 0 & 0 & 0 & 0 & 0 & 0 & 0 & 0 & 0 & 0 & 0 & 0 & 0 & 0 \\
 0 & 0 & 0 & 0 & 0 & 0 & 0 & 0 & 0 & 0 & 0 & 0 & 0 & 0 & 0 & \six & 0 & 0 & 0 & 0 & 0 & 0 & 0 & 0 & 0 & 0 & 0 & 0 & 0 & 0 & 0 & 0 & 0 & 0 & 0 & 0 & 0 & 0 & 0 & 0 & 0 & 0 & 0 & 0 & 0 & 0 & 0 & 0 & 0 & 0 & 0 \\
 0 & 0 & 0 & 0 & 0 & 0 & 0 & 0 & 0 & 0 & 0 & 0 & 0 & 0 & 0 & 0 & \six & 0 & 0 & 0 & 0 & 0 & 0 & 0 & 0 & 0 & 0 & 0 & 0 & 0 & 0 & 0 & 0 & 0 & 0 & 0 & 0 & 0 & 0 & 0 & 0 & 0 & 0 & 0 & 0 & 0 & 0 & 0 & 0 & 0 & 0 \\
 0 & 0 & 0 & 0 & 0 & 0 & 0 & 0 & 0 & 0 & 0 & 0 & 0 & 0 & 0 & 0 & 0 & \six & 0 & 0 & 0 & 0 & 0 & 0 & 0 & 0 & 0 & 0 & 0 & 0 & 0 & 0 & 0 & 0 & 0 & 0 & 0 & 0 & 0 & 0 & 0 & 0 & 0 & 0 & 0 & 0 & 0 & 0 & 0 & 0 & 0 \\
 0 & 0 & 0 & 0 & 0 & 0 & 0 & 0 & 0 & 0 & 0 & 0 & 0 & 0 & 0 & 0 & 0 & 0 & \six & 0 & 0 & 0 & 0 & 0 & 0 & 0 & 0 & 0 & 0 & 0 & 0 & 0 & 0 & 0 & 0 & 0 & 0 & 0 & 0 & 0 & 0 & 0 & 0 & 0 & 0 & 0 & 0 & 0 & 0 & 0 & 0 \\
 0 & 0 & 0 & 0 & 0 & 0 & 0 & 0 & 0 & 0 & 0 & 0 & 0 & 0 & 0 & 0 & 0 & 0 & 0 & \six & 0 & 0 & 0 & 0 & 0 & 0 & 0 & 0 & 0 & 0 & 0 & 0 & 0 & 0 & 0 & 0 & 0 & 0 & 0 & 0 & 0 & 0 & 0 & 0 & 0 & 0 & 0 & 0 & 0 & 0 & 0 \\
 0 & 0 & 0 & 0 & 0 & 0 & 0 & 0 & 0 & 0 & 0 & 0 & 0 & 0 & 0 & 0 & 0 & 0 & 0 & 0 & \six & 0 & 0 & 0 & 0 & 0 & 0 & 0 & 0 & 0 & 0 & 0 & 0 & 0 & 0 & 0 & 0 & 0 & 0 & 0 & 0 & 0 & 0 & 0 & 0 & 0 & 0 & 0 & 0 & 0 & 0 \\
 0 & 0 & 0 & 0 & 0 & 0 & 0 & 0 & 0 & 0 & 0 & 0 & 0 & 0 & 0 & 0 & 0 & 0 & 0 & 0 & 0 & \six & 0 & 0 & 0 & 0 & 0 & 0 & 0 & 0 & 0 & 0 & 0 & 0 & 0 & 0 & 0 & 0 & 0 & 0 & 0 & 0 & 0 & 0 & 0 & 0 & 0 & 0 & 0 & 0 & 0 \\
 0 & 0 & 0 & 0 & 0 & 0 & 0 & 0 & 0 & 0 & 0 & 0 & 0 & 0 & 0 & 0 & 0 & 0 & 0 & 0 & 0 & 0 & \six & 0 & 0 & 0 & 0 & 0 & 0 & 0 & 0 & 0 & 0 & 0 & 0 & 0 & 0 & 0 & 0 & 0 & 0 & 0 & 0 & 0 & 0 & 0 & 0 & 0 & 0 & 0 & 0 \\
 0 & 0 & 0 & 0 & 0 & 0 & 0 & 0 & 0 & 0 & 0 & 0 & 0 & 0 & 0 & 0 & 0 & 0 & 0 & 0 & 0 & 0 & 0 & \six & 0 & 0 & 0 & 0 & 0 & 0 & 0 & 0 & 0 & 0 & 0 & 0 & 0 & 0 & 0 & 0 & 0 & 0 & 0 & 0 & 0 & 0 & 0 & 0 & 0 & 0 & 0 \\
 0 & 0 & 0 & 0 & 0 & 0 & 0 & 0 & 0 & 0 & 0 & 0 & 0 & 0 & 0 & 0 & 0 & 0 & 0 & 0 & 0 & 0 & 0 & 0 & \six & 0 & 0 & 0 & 0 & 0 & 0 & 0 & 0 & 0 & 0 & 0 & 0 & 0 & 0 & 0 & 0 & 0 & 0 & 0 & 0 & 0 & 0 & 0 & 0 & 0 & 0 \\
 0 & 0 & 0 & 0 & 0 & 0 & 0 & 0 & 0 & 0 & 0 & 0 & 0 & 0 & 0 & 0 & 0 & 0 & 0 & 0 & 0 & 0 & 0 & 0 & 0 & \six & 0 & 0 & 0 & 0 & 0 & 0 & 0 & 0 & 0 & 0 & 0 & 0 & 0 & 0 & 0 & 0 & 0 & 0 & 0 & 0 & 0 & 0 & 0 & 0 & 0 \\
 0 & 0 & 0 & 0 & 0 & 0 & 0 & 0 & 0 & 0 & 0 & 0 & 0 & 0 & 0 & 0 & 0 & 0 & 0 & 0 & 0 & 0 & 0 & 0 & 0 & 0 & \six & 0 & 0 & 0 & 0 & 0 & 0 & 0 & 0 & 0 & 0 & 0 & 0 & 0 & 0 & 0 & 0 & 0 & 0 & 0 & 0 & 0 & 0 & 0 & 0 \\
 0 & 0 & 0 & 0 & 0 & 0 & 0 & 0 & 0 & 0 & 0 & 0 & 0 & 0 & 0 & 0 & 0 & 0 & 0 & 0 & 0 & 0 & 0 & 0 & 0 & 0 & 0 & \six & 0 & 0 & 0 & 0 & 0 & 0 & 0 & 0 & 0 & 0 & 0 & 0 & 0 & 0 & 0 & 0 & 0 & 0 & 0 & 0 & 0 & 0 & 0 \\
 0 & 0 & 0 & 0 & 0 & 0 & 0 & 0 & 0 & 0 & 0 & 0 & 0 & 0 & 0 & 0 & 0 & 0 & 0 & 0 & 0 & 0 & 0 & 0 & 0 & 0 & 0 & 0 & \six & 0 & 0 & 0 & 0 & 0 & 0 & 0 & 0 & 0 & 0 & 0 & 0 & 0 & 0 & 0 & 0 & 0 & 0 & 0 & 0 & 0 & 0 \\
 0 & 0 & 0 & 0 & 0 & 0 & 0 & 0 & 0 & 0 & 0 & 0 & 0 & 0 & 0 & 0 & 0 & 0 & 0 & 0 & 0 & 0 & 0 & 0 & 0 & 0 & 0 & 0 & 0 & \six & 0 & 0 & 0 & 0 & 0 & 0 & 0 & 0 & 0 & 0 & 0 & 0 & 0 & 0 & 0 & 0 & 0 & 0 & 0 & 0 & 0 \\
 0 & 0 & 0 & 0 & 0 & 0 & 0 & 0 & 0 & 0 & 0 & 0 & 0 & 0 & 0 & 0 & 0 & 0 & 0 & 0 & 0 & 0 & 0 & 0 & 0 & 0 & 0 & 0 & 0 & 0 & \six & 0 & 0 & 0 & 0 & 0 & 0 & 0 & 0 & 0 & 0 & 0 & 0 & 0 & 0 & 0 & 0 & 0 & 0 & 0 & 0 \\
 0 & 0 & 0 & 0 & 0 & 0 & 0 & 0 & 0 & 0 & 0 & 0 & 0 & 0 & 0 & 0 & 0 & 0 & 0 & 0 & 0 & 0 & 0 & 0 & 0 & 0 & 0 & 0 & 0 & 0 & 0 & \six & 0 & 0 & 0 & 0 & 0 & 0 & 0 & 0 & 0 & 0 & 0 & 0 & 0 & 0 & 0 & 0 & 0 & 0 & 0 \\
 0 & 0 & 0 & 0 & 0 & 0 & 0 & 0 & 0 & 0 & 0 & 0 & 0 & 0 & 0 & 0 & 0 & 0 & 0 & 0 & 0 & 0 & 0 & 0 & 0 & 0 & 0 & 0 & 0 & 0 & 0 & 0 & \six & 0 & 0 & 0 & 0 & 0 & 0 & 0 & 0 & 0 & 0 & 0 & 0 & 0 & 0 & 0 & 0 & 0 & 0 \\
 0 & 0 & 0 & 0 & 0 & 0 & 0 & 0 & 0 & 0 & 0 & 0 & 0 & 0 & 0 & 0 & 0 & 0 & 0 & 0 & 0 & 0 & 0 & 0 & 0 & 0 & 0 & 0 & 0 & 0 & 0 & 0 & 0 & \six & 0 & 0 & 0 & 0 & 0 & 0 & 0 & 0 & 0 & 0 & 0 & 0 & 0 & 0 & 0 & 0 & 0 \\
 0 & 0 & 0 & 0 & 0 & 0 & 0 & 0 & 0 & 0 & 0 & 0 & 0 & 0 & 0 & 0 & 0 & 0 & 0 & 0 & 0 & 0 & 0 & 0 & 0 & 0 & 0 & 0 & 0 & 0 & 0 & 0 & 0 & 0 & \six & 0 & 0 & 0 & 0 & 0 & 0 & 0 & 0 & 0 & 0 & 0 & 0 & 0 & 0 & 0 & 0 \\
 0 & 0 & 0 & 0 & 0 & 0 & 0 & 0 & 0 & 0 & 0 & 0 & 0 & 0 & 0 & 0 & 0 & 0 & 0 & 0 & 0 & 0 & 0 & 0 & 0 & 0 & 0 & 0 & 0 & 0 & 0 & 0 & 0 & 0 & 0 & \six & 0 & 0 & 0 & 0 & 0 & 0 & 0 & 0 & 0 & 0 & 0 & 0 & 0 & 0 & 0 \\
 0 & 0 & 0 & 0 & 0 & 0 & 0 & 0 & 0 & 0 & 0 & 0 & 0 & 0 & 0 & 0 & 0 & 0 & 0 & 0 & 0 & 0 & 0 & 0 & 0 & 0 & 0 & 0 & 0 & 0 & 0 & 0 & 0 & 0 & 0 & 0 & \six & 0 & 0 & 0 & 0 & 0 & 0 & 0 & 0 & 0 & 0 & 0 & 0 & 0 & 0 \\
 0 & 0 & 0 & 0 & 0 & 0 & 0 & 0 & 0 & 0 & 0 & 0 & 0 & 0 & 0 & 0 & 0 & 0 & 0 & 0 & 0 & 0 & 0 & 0 & 0 & 0 & 0 & 0 & 0 & 0 & 0 & 0 & 0 & 0 & 0 & 0 & 0 & \six & 0 & 0 & 0 & 0 & 0 & 0 & 0 & 0 & 0 & 0 & 0 & 0 & 0 \\
 0 & 0 & 0 & 0 & 0 & 0 & 0 & 0 & 0 & 0 & 0 & 0 & 0 & 0 & 0 & 0 & 0 & 0 & 0 & 0 & 0 & 0 & 0 & 0 & 0 & 0 & 0 & 0 & 0 & 0 & 0 & 0 & 0 & 0 & 0 & 0 & 0 & 0 & \six & 0 & 0 & 0 & 0 & 0 & 0 & 0 & 0 & 0 & 0 & 0 & 0 \\
 0 & 0 & 0 & 0 & 0 & 0 & 0 & 0 & 0 & 0 & 0 & 0 & 0 & 0 & 0 & 0 & 0 & 0 & 0 & 0 & 0 & 0 & 0 & 0 & 0 & 0 & 0 & 0 & 0 & 0 & 0 & 0 & 0 & 0 & 0 & 0 & 0 & 0 & 0 & \six & 0 & 0 & 0 & 0 & 0 & 0 & 0 & 0 & 0 & 0 & 0 \\
 0 & 0 & 0 & 0 & 0 & 0 & 0 & 0 & 0 & 0 & 0 & 0 & 0 & 0 & 0 & 0 & 0 & 0 & 0 & 0 & 0 & 0 & 0 & 0 & 0 & 0 & 0 & 0 & 0 & 0 & 0 & 0 & 0 & 0 & 0 & 0 & 0 & 0 & 0 & 0 & \six & 0 & 0 & 0 & 0 & 0 & 0 & 0 & 0 & 0 & 0 \\
 0 & 0 & 0 & 0 & 0 & 0 & 0 & 0 & 0 & 0 & 0 & 0 & 0 & 0 & 0 & 0 & 0 & 0 & 0 & 0 & 0 & 0 & 0 & 0 & 0 & 0 & 0 & 0 & 0 & 0 & 0 & 0 & 0 & 0 & 0 & 0 & 0 & 0 & 0 & 0 & 0 & \six & 0 & 0 & 0 & 0 & 0 & 0 & 0 & 0 & 0 \\
 0 & 0 & 0 & 0 & 0 & 0 & 0 & 0 & 0 & 0 & 0 & 0 & 0 & 0 & 0 & 0 & 0 & 0 & 0 & 0 & 0 & 0 & 0 & 0 & 0 & 0 & 0 & 0 & 0 & 0 & 0 & 0 & 0 & 0 & 0 & 0 & 0 & 0 & 0 & 0 & 0 & 0 & \six & 0 & 0 & 0 & 0 & 0 & 0 & 0 & 0 \\
 0 & 0 & 0 & 0 & 0 & 0 & 0 & 0 & 0 & 0 & 0 & 0 & 0 & 0 & 0 & 0 & 0 & 0 & 0 & 0 & 0 & 0 & 0 & 0 & 0 & 0 & 0 & 0 & 0 & 0 & 0 & 0 & 0 & 0 & 0 & 0 & 0 & 0 & 0 & 0 & 0 & 0 & 0 & \six & 0 & 0 & 0 & 0 & 0 & 0 & 0 \\
 0 & 0 & 0 & 0 & 0 & 0 & 0 & 0 & 0 & 0 & 0 & 0 & 0 & 0 & 0 & 0 & 0 & 0 & 0 & 0 & 0 & 0 & 0 & 0 & 0 & 0 & 0 & 0 & 0 & 0 & 0 & 0 & 0 & 0 & 0 & 0 & 0 & 0 & 0 & 0 & 0 & 0 & 0 & 0 & \six & 0 & 0 & 0 & 0 & 0 & 0 \\
 0 & 0 & 0 & 0 & 0 & 0 & 0 & 0 & 0 & 0 & 0 & 0 & 0 & 0 & 0 & 0 & 0 & 0 & 0 & 0 & 0 & 0 & 0 & 0 & 0 & 0 & 0 & 0 & 0 & 0 & 0 & 0 & 0 & 0 & 0 & 0 & 0 & 0 & 0 & 0 & 0 & 0 & 0 & 0 & 0 & \six & 0 & 0 & 0 & 0 & 0 \\
 0 & 0 & 0 & 0 & 0 & 0 & 0 & 0 & 0 & 0 & 0 & 0 & 0 & 0 & 0 & 0 & 0 & 0 & 0 & 0 & 0 & 0 & 0 & 0 & 0 & 0 & 0 & 0 & 0 & 0 & 0 & 0 & 0 & 0 & 0 & 0 & 0 & 0 & 0 & 0 & 0 & 0 & 0 & 0 & 0 & 0 & \six & 0 & 0 & 0 & 0 \\
 0 & 0 & 0 & 0 & 0 & 0 & 0 & 0 & 0 & 0 & 0 & 0 & 0 & 0 & 0 & 0 & 0 & 0 & 0 & 0 & 0 & 0 & 0 & 0 & 0 & 0 & 0 & 0 & 0 & 0 & 0 & 0 & 0 & 0 & 0 & 0 & 0 & 0 & 0 & 0 & 0 & 0 & 0 & 0 & 0 & 0 & 0 & \six & 0 & 0 & 0 \\
 0 & 0 & 0 & 0 & 0 & 0 & 0 & 0 & 0 & 0 & 0 & 0 & 0 & 0 & 0 & 0 & 0 & 0 & 0 & 0 & 0 & 0 & 0 & 0 & 0 & 0 & 0 & 0 & 0 & 0 & 0 & 0 & 0 & 0 & 0 & 0 & 0 & 0 & 0 & 0 & 0 & 0 & 0 & 0 & 0 & 0 & 0 & 0 & \six & 0 & 0 \\
 0 & 0 & 0 & 0 & 0 & 0 & 0 & 0 & 0 & 0 & 0 & 0 & 0 & 0 & 0 & 0 & 0 & 0 & 0 & 0 & 0 & 0 & 0 & 0 & 0 & 0 & 0 & 0 & 0 & 0 & 0 & 0 & 0 & 0 & 0 & 0 & 0 & 0 & 0 & 0 & 0 & 0 & 0 & 0 & 0 & 0 & 0 & 0 & 0 & \six & 0 \\
 0 & 0 & 0 & 0 & 0 & 0 & 0 & 0 & 0 & 0 & 0 & 0 & 0 & 0 & 0 & 0 & 0 & 0 & 0 & 0 & 0 & 0 & 0 & 0 & 0 & 0 & 0 & 0 & 0 & 0 & 0 & 0 & 0 & 0 & 0 & 0 & 0 & 0 & 0 & 0 & 0 & 0 & 0 & 0 & 0 & 0 & 0 & 0 & 0 & 0 & \six \\
\end{array}
\right)}\,.&
\end{flalign}

\subsection{Field space diameter}
We are now in a position to determine the diameter of the fundamental domain in the DDFGK compactification. The diameter is given by (\ref{dynrange}),
\be
{\cal D}_{\text{light}}=2\pi |\diag{\xi_i}\,{\mathbf S}_\Xi^\top \resc{\hat{\mathbf v}}|\, ,
\ee
where $\hat{\mathbf v}=S_{\Xi}\,\diag(\xi^{-1}) \Psi^{M^\Phi}_1$. Using the Hessian matrix for the axions in (\ref{eqn:axionsaxionhess}), the $\mathbf q$ matrix (\ref{fullq}), and the K\"ahler  metric on moduli space, we numerically find that the diameter along the lightest direction\footnote{Note that by using different choices of $q$, corresponding to different coordinates, the eigenvalues of the kinetic matrix change. We have observed examples in which $\xi_{N}\approx 16\M$, which  might naively be interpreted as a super-Planckian axion decay constant. However, as the definition of the axions is ambiguous in this example, this does {\it not} correspond to a physically large diameter.} is
\be\label{ddfgkrange}
{\cal D}_{\text{light}}=1.13\,\M\, .
\ee

This can be compared to the results of \S\ref{sec:align}, where the field space diameter  was obtained analytically. As we argued in \S\ref{sec:align}, it is  reasonable to approximate the kinetic matrix $\mathbf \Xi$ as an inverse Wishart matrix. We can test this assumption by comparing the largest eigenvalue of the kinetic matrix obtained from the K\"ahler potential (\ref{kahlerpot}) to the typical largest eigenvalue of a Wishart matrix given in (\ref{wismaxev}). Using the scale of the $\mathbf q$ matrix (\ref{fullq}), $\sigma_{\cal Q} \approx 0.18$, and the largest metric eigenvalue $f_{N}\approx0.013\,\M$, (\ref{wismaxev}) gives $\xi_N^{\text{Wishart}}\approx 0.49\,\M$, while numerically we typically\footnote{Note again that the kinetic matrix is basis dependent. We obtained a typical value by evaluating $\xi_N$ for a large number of random basis choices.} find $\xi_N^{\text{DDFGK}}\approx 0.18\,\M$. According to (\ref{wisrange}), the field space diameter is obtained by rescaling the largest eigenvalue of the kinetic matrix by $\lVert \resc{\hat{\mathbf v}}\rVert$, which takes into account the additional $P-N=9$ constraints. From (\ref{qrectangulardiam})  we expect that for random choices of constraints, $\lVert \resc{\hat{\mathbf v}}\rVert\approx 2.5$, while we observe numerically that for the direction $\hat{\mathbf v}$ corresponding to the lightest canonically normalized field, $\lVert \resc{\hat{\mathbf v}}\rVert\approx 6.3$.    It is encouraging that our large $N$ estimates based on universality and eigenvector delocalization are accurate, in this example, to within factors of order a few.

Finally, (\ref{wisrange}) gives an analytic estimate for the field space diameter from random matrix theory of
\be
{\cal D}=\sqrt{\frac{51}{2+\log (4)-2 \sqrt{1+\log (4)}}}    \frac{\pi }{2 \,\text{erf}^{-1}\left(2^{-1/9}\right)}   {f_{N}\over \sigma_{\cal Q}} \approx1.21\,\M\, .
\ee
This matches the actual diameter (\ref{ddfgkrange}) rather well.

\section{A Unified Theory of Axion Diameters} \label{unifiedtheory}

Our results unify a number of effects identified in prior works, as we will now explain.\footnote{For simplicity of presentation we take $P=N$ in this discussion.}
The very special case ${\mathbf K}=\diag(f_i^2)$, ${\mathbf Q}=\mathbb{1}$ corresponds to the simplest construction of N-flation \cite{Nflation} (a version of assisted inflation \cite{Liddle:1998jc}), for which the field range is  given by the Pythagorean sum $\Delta\Phi=2\pi\sqrt{\sum_i f_i^2}$.
In the much more general circumstance where ${\mathbf K}$  is not diagonal in the basis where ${\mathbf Q}=\mathbb{1}$, eigenvector delocalization causes the eigenvector $\Psi_N^{K}$  with the largest eigenvalue $f_N^2$  to point in an approximately diagonal direction, leading to the range $\Delta\Phi=2\pi \sqrt{N} f_N$ \cite{Bachlechner:2014hsa}.
The result of the present work is closely parallel to that of \cite{Bachlechner:2014hsa}: we have seen that when
${\mathbf \Xi}=({\mathbf Q}^{-1})^\top {\mathbf K}\, {\mathbf Q}^{-1}$ is not diagonal in the basis where ${\mathbf Q}=\mathbb{1}$, eigenvector delocalization causes the eigenvector $\Psi_N^{\Xi}$  with the largest eigenvalue $\xi_N^2$ to point in an approximately diagonal direction, leading to the range $\Delta\Phi=2\pi \sqrt{N} \xi_N$.

To understand the crucial distinction between $\xi_N$ and $f_N$,  it is useful to work in the concrete case of Calabi-Yau compactifications of type IIB string theory.
In this setting we notice that ${\mathbf K}$ can be computed in terms of classical data, namely the intersection numbers. At this level, the axion field space is $\mathbb{R}^N$; the axions have vanishing potential, have infinite range, and do not decay.   Meaningful statements about axion decay constants require specifying the nonperturbative effects that break the continuous shift symmetries to discrete shifts, which are encoded in ${\mathbf Q}$.   For this reason, for any $N>1$, a computation of the eigenvalues $f_i^2$  of the K\"ahler metric ${\mathbf K}$  defined by the classical K\"ahler potential {\it{does not}}  determine the physical field range.\footnote{The bound on the diameter of axion moduli space obtained for simplicial K\"ahler cones in \cite{Rudelius:2014wla} uses only the data of ${\mathbf K}$, taking ${\mathbf Q}=\mathbb{1}$, and does not apply in the general case where ${\mathbf Q}\neq\mathbb{1}$.}  In particular, an upper bound on $f_N$ does not provide an upper bound on the possible axion displacement during inflation, for two reasons.   First, $\Delta\Phi/f_N$  is parametrically large at large $N$ --- as large as ${\cal O}(N^{3/2})$ --- for generic ${\mathbf K}$ and ${\mathbf Q}$.  Second, even for $N=2$, there is the possibility that the smallest eigenvalue $\lambda_1^{Q^\top Q}$ of $\mathbf Q^\top \mathbf Q$ is accidentally small in comparison to its expected size $\langle\lambda_1^{Q^\top Q}\rangle$  in an ensemble of $\mathbf Q$  matrices with the same symmetries and with entries of the same r.m.s.~size.

The possibility that $\lambda_1^{Q^\top Q} \ll \langle\lambda_1^{Q^\top Q}\rangle$ is the foundation of the Kim-Nilles-Peloso (KNP) mechanism of decay constant alignment \cite{KNP}.
The proposal of KNP, described for $N=2$ in \cite{KNP} and generalized to $N>2$ in \cite{Choi:2014rja}, is to take ${\mathbf K}=\diag(f_i^2)$ in a basis where ${\mathbf Q}$  is nontrivial, and to take $\mathbf Q^\top \mathbf Q$ to have an accidentally small smallest eigenvalue.  Such an accidental enhancement is plausibly realizable in the landscape of string vacua, but for $N=2$ ---  and indeed for any $N$  that is not large ---  this occurs infrequently \cite{HT1}.  The increased likelihood at large $N$ of large enhancements from small $\lambda_1^{Q^\top Q}/\langle\lambda_1^{Q^\top Q}\rangle$ was observed by Higaki and Takahashi in \cite{HT1} (see also \cite{HT2}), and a slightly different perspective on enhancements at large N, also building on \cite{KNP}, was given by Choi, Kim, and Yun in \cite{Choi:2014rja}.

Here we have not relied on $\lambda_1^{Q^\top Q} \ll \langle\lambda_1^{Q^\top Q}\rangle$, but have instead shown that for ${\mathbf Q}$ matrices of the form that arise in actual string compactifications, $\langle\lambda_1^{Q^\top Q}\rangle$ itself is small, because of eigenvalue repulsion.   Thus, the field range computed  in this work is the {\it{the generic circumstance}}, not a fine-tuned possibility.\footnote{The range we have exhibited is an `enhancement' compared to prior expectations, but it would be more accurate to say that those prior works that considered only the $f_i$, rather than the $\xi_i$,  underestimated the typical diameter of field space.}

A potential obstruction to achieving a super-Planckian displacement in a theory with an extremely large number of axions is that
renormalization of the Planck mass (cf.~\cite{Nflation}) reduces the effective range $\Delta\Phi$, measured in renormalized Planck mass units.
General reasoning suggests
\begin{equation}  \label{naiverenormalization}
(\M^{{\rm ren.}})^2 - (\M^{{\rm bare}})^2 \equiv \delta \M^2 \sim \frac{N}{16\pi^2}\Lambda_{\rm{UV}}^2\,,
\end{equation} where $\Lambda_{\rm{UV}}$ is the ultraviolet cutoff.  However, (\ref{naiverenormalization}) is manifestly ultraviolet sensitive, and a more meaningful approach is to examine  the leading correction that arises in string theory.   Compactifying type IIB string theory on a six-manifold $X_6$ with Euler characteristic $\chi(X_6)$ and volume ${\cal{V}}$,  and including  the four-loop $\sigma$-model coupling quartic in ten-dimensional curvature~\cite{Gross:1986iv, Becker:2002nn}, one finds
\begin{equation}  \label{sigmarenormalization}
\delta \M^2 = \M^2 \times \frac{\zeta(3)\chi(X_6)}{8(2\pi)^3 g_s^{3/2} }\frac{(l_s)^6}{{\cal{V}}}\,,
\end{equation}
where $\cal{V}$ is the Einstein frame volume of the orientifold, and we are using the conventions of~\cite{Denef:2005mm}. If the axions in question arise in the K\"ahler moduli sector, so that $N=h^{1,1}$,  the correction (\ref{sigmarenormalization})  has the same parametric scaling as (\ref{naiverenormalization}), if $h^{1,1}$  is taken large with $h^{2,1}$  fixed.   However, in typical Calabi-Yau compactifications, (\ref{sigmarenormalization}) is a modest correction, $\delta \M^2 \lesssim \M^2$, and does not parametrically alter the field range.   In the example of DDFGK, $\delta \M^2/\M^2 = 0.008$.  We conclude that renormalization of the Planck mass does not present a serious obstacle to achieving super-Planckian axion diameters in reasonable Calabi-Yau compactifications through our approach, though it would become problematic at the very large values of $N$  needed in N-flation models \cite{Nflation} with ${\mathbf Q} = \mathbb{1}$.

\section{Conclusions}\label{sec:conclusion}

We have computed the diameter ${\cal D}$ of the axion fundamental domain in a general field theory with $N$ axions, with the Lagrangian
\be
{\mathcal L}={1\over 2}K_{ij} \partial\theta^i\partial\theta^j - \sum_{i=1}^N \Lambda_i^4 \left[1-\cos\left({\cal Q}^i_{\, j}\theta^j \right) \right]\,,
\ee where ${\cal Q}$ is a $P \times N$ matrix of integers defining the periodic identifications of the axions.
One key result is the diameter (\ref{kinrange}) along a particular direction, which gives a deterministic lower bound on the maximal diameter.
We evaluated (\ref{kinrange}) in various regimes using results from random matrix theory, leading to approximate lower bounds that hold with high confidence at large $N$.
The resulting scalings with $N$ are shown in Table \ref{table123}.
\begin{table}\centering
\ra{1.3}
\begin{tabular}{@{}p{2.5cm}p{1.5cm}p{1.5cm}cp{1.5cm}p{1.5cm}@{}}\toprule
& \multicolumn{2}{c}{${\cal Q}^\top {\cal Q}~~(P=N)$} & \phantom{abc}&  \multicolumn{2}{c}{${\cal Q}^\top {\cal Q}~~(P>N)$}\\ \cmidrule{2-3} \cmidrule{5-6} \raggedright ${\mathbf K}$& \raggedleft Unit\phantom{i} &\raggedleft Wishart  && \multicolumn{2}{c}{ Wishart} \\ \midrule
 \raggedright Unit & \raggedleft$ \sqrt{N} f $&\raggedleft $ N^{3/2}f $ &&\multicolumn{2}{r}{$N f $\phantom{abci}} \\
 \raggedright Wishart & \raggedleft$ \sqrt{N} f_{N} $&\raggedleft$N^{3/2} f_{N} $&&\multicolumn{2}{r}{$Nf_{N} $\phantom{abci}}\\
 \raggedright Heavy Tailed & \raggedleft $f_{N} $&\raggedleft $\phantom{Jo} N f_{N} $&&\multicolumn{2}{r}{$\sqrt{N}f_{N}$\phantom{abci}} \\ \bottomrule
\end{tabular}
\caption{\label{table123}Parametric scaling of the maximum diameter of the axion fundamental domain for different choices of metrics $\mathbf K$ and axion constraints $\mathcal Q$. $P$ is the number of constraints,  $N$ is the number of axion fields,  and $f_N^2$  is the largest eigenvalue of $\mathbf K$.}
\end{table}

We substantiated our general findings  by computing the diameter of the axion fundamental domain in explicit Calabi-Yau compactifications of string theory.
We focused on the vacuum of F-theory constructed in \cite{Denef:2005mm}, where all moduli are fixed in a regime where known higher-order corrections are controllably small.
The nonperturbative superpotential generated by Euclidean D3-branes and by gaugino condensation on D7-branes defines a specific $51\times 60$ ${\cal Q}$ matrix (\ref{fullq}) for the $h^{1,1}=51$ Ramond-Ramond axions that complexify the K\"ahler moduli.
For the precise vacuum parameters taken in \cite{Denef:2005mm}, where higher order corrections are parametrically controlled, the largest metric eigenvalue obeys $f_N \approx 0.013 \M$.  Our random matrix results predict ${\cal D} \gtrsim \M$, and by direct computation we have confirmed that ${\cal D} \gtrsim 1.1 \M$.

Let us close by discussing the potential implications of our results. There are a number of arguments against the possibility of arbitrarily large displacements $\Delta\Phi$ of scalar fields in effective theories that admit completions in quantum gravity.\footnote{See e.g.~\cite{Banks:2003sx,Ooguri:2006in}, as well as the recent review \cite{Liamsbook}.}   However, it has proved difficult to sharpen general quantum gravity arguments to place accurate limits $\Delta\Phi<\M$, as contrasted with $\Delta\Phi<\infty$: the maximal $\Delta\Phi$ in a given theory depends on the details of the ultraviolet completion, and existing general arguments are not precise enough to capture factors of order $\pi$.  Moreover, there are mechanisms implying the plausible existence of counterexamples --- constructions of large-field inflation in string theory --- based on effects such as decay constant alignment \cite{KNP}, N-flation \cite{Nflation}, or monodromy \cite{SW,MSW}.  These proposals have not yet led to universally acknowledged existence proofs of large-field inflation in string theory, because of the difficulty of embedding these mechanisms
into explicit and parametrically controlled compactifications  with stabilized moduli.

Our findings present a way forward: they provide a framework for exhibiting super-Planckian axion displacements in well-understood vacua of string theory, without fine-tuning of parameters, and without working at extremely large $N \gtrsim 10^3$.  By unifying the decay constant alignment effect of KNP \cite{KNP} with the eigenvector delocalization described in \cite{Bachlechner:2014hsa}, and arguing that both effects are generically present, we have shown that the diameter of axion field space is parametrically larger in $N \gg 1$  than was anticipated in the context of N-flation \cite{Nflation,Bachlechner:2014hsa}.  Our results hold in a broad class of theories in which ${\cal Q}$ is a somewhat sparse matrix, and we argued that many flux compactifications on Calabi-Yau orientifolds fall into this category.  While our field theoretic arguments apply for any $N \gg 1$, in this work the largest number of axions we have examined in an explicit vacuum of string theory is $N=h^{1,1}=51$, in the case of the DDFGK compactification of F-theory \cite{Denef:2005mm}.  Because ${\cal D} \approx \M$ in this example, we anticipate that displacements suitable for large-field inflation, $\Delta\Phi \gtrsim 10 \M$,  could be achieved in a compactification with similar structures but with $h^{1,1}$ of order a few hundred, comfortably inside the range of known Calabi-Yau threefolds.  Exhibiting an example of this sort is an important problem for the future.

We have argued that in a theory consisting solely of $N$ axions, inflationary evolution can rather naturally proceed along the super-Planckian diameters that we have identified.
However, in compactifications with spontaneously broken supersymmetry, including the example of \cite{Denef:2005mm}, the couplings of saxions to axions may lead to instabilities that preclude inflation.  This is a general difficulty: even in vacua of string theory that admit super-Planckian axion displacements, the  uncontrolled evolution of moduli fields presents a challenge  for any candidate construction of large-field inflation.   The theories we have described here are a promising arena for grappling with this fundamental problem.

\section*{Acknowledgements}
We thank  Raphael Bousso, Thorsten Rahn, John Stout, and Timm Wrase for useful discussions, and we thank Diederik Roest and Cliff Burgess for sharing their related results with us.
This work was supported by NSF grant PHY-0757868 and by a Simons Fellowship.

\appendix

\section{\label{app:RMT}Results from RMT}
In the study of theories with $N \gg 1$  scalar fields, relevant matrix quantities such as the metric on field space and the Hessian matrix approach a universal limit that is governed by random matrix theory. This emergent behavior  is a powerful tool for studying random supergravity theories \cite{Marsh:2011aa,Bachlechner:2012at,Bachlechner:2014rqa}. In this section we review a few results from random matrix theory  that are needed in this work. A more comprehensive review of random matrix theory and its application in physics can be found in \cite{dumitriu1,mehta2004random,CambridgeJournals:298726}.

\subsection{Classical ensembles}
Random matrix ensembles  can be classified by their symmetry properties. Two classes of physical relevance are the Hermite (Wigner) and Laguerre (Wishart) $\beta$-ensembles. Consider a random $N\times N$ matrix $\mathbf A$ with entries that are independent, identically distributed (i.i.d.) random numbers of variance $\sigma^2$. The ensemble of Wigner matrices with $\beta=1,2$ are defined by
\be
M_H=\mathbf A+\mathbf A^\dagger\, ,
\ee
while the Wishart ensemble is defined in terms of an $M\times N$ matrix $\mathbf A$
\be
M_L=\mathbf A\cdot \mathbf A^\dagger\, ,
\ee
where $\beta=1$ corresponds to real entries in $\mathbf A$, while $\beta=2$ corresponds to complex entries. These are rotationally invariant ensembles of random matrices.
In the large $N$ limit, the precise probability distribution for the entries of $\mathbf A$ loses relevance (as long as its variance is sufficiently bounded), and a universal limit is approached. In this limit, the symmetry properties of the ensemble define statistical observables such as the eigenvalue and eigenvector distributions. Table \ref{rmtt} lists some properties of the Wigner and Wishart ensembles in large $N$ limit \cite{dumitriu1}.

\begin{table}\begin{center}
 \includegraphics[width=.96\textwidth]{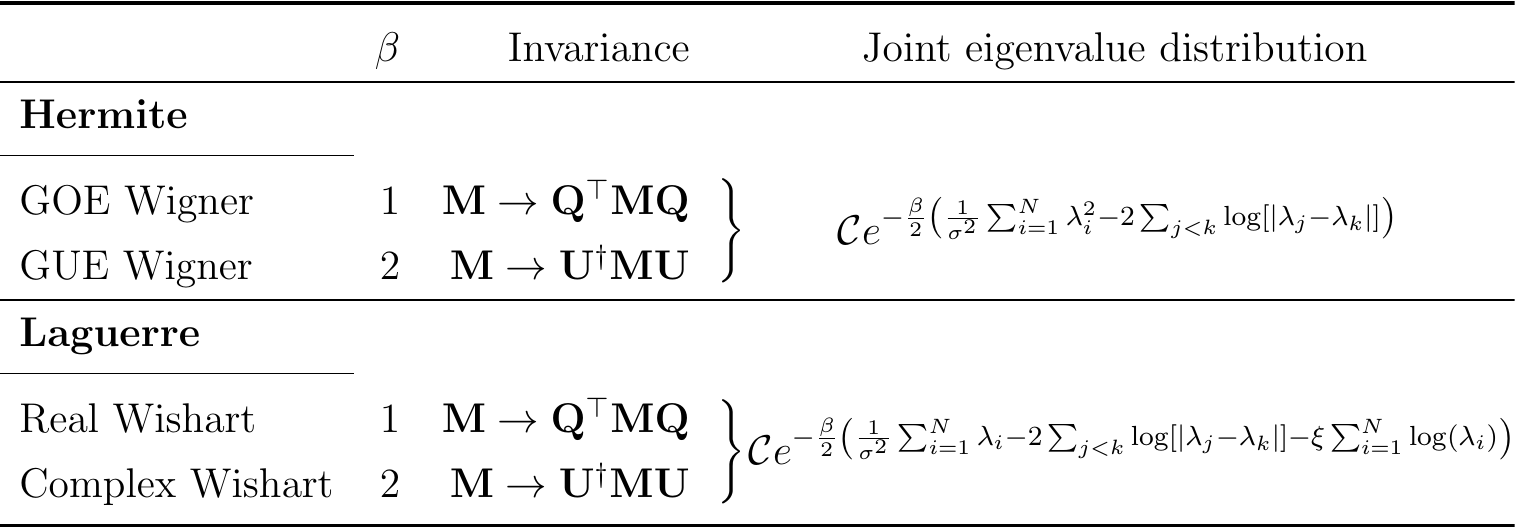}
 \end{center}
\caption{Summary of Hermite and Laguerre random matrix ensembles \cite{dumitriu1}. The matrix $\mathbf Q$ represents an orthogonal transformation, while $\mathbf U$ represents a unitary transformation. In the joint eigenvalue distribution, the constant $\xi$ is given by $\xi=M-N+1-2/\beta$.\label{rmtt}}
\end{table}

Note in particular that the joint eigenvalue distribution of both the Wigner and the Wishart ensemble can be interpreted as the probability distributions of a classical, one-dimensional gas at finite temperature $1/\beta$ with Coulomb interactions. The probability is given by $\rho(\lambda_i)=e^{\beta H}$.
In the large $N$ limit, the eigenvalue spectrum of the Wigner ensemble is given by the Wigner semicircle law,
\be
\rho(\lambda)={1\over 2\pi N\sigma^2}\sqrt{4N \sigma^2-\lambda^2}\,,
\ee
while the eigenvalue spectrum of the square Wishart ensemble is given by
\be
\rho(\lambda)={1\over 2\pi N\sigma^2\lambda}\sqrt{(4 N\sigma^2-\lambda)\lambda}\,.
\ee
The typical scale of the largest eigenvalue in the large $N$ limit is given by $\lambda_{N}^{{\rm{Wis}}}= 4\sigma^2 N$.
By interpreting the random matrix ensembles  in terms of an interacting gas at finite temperature with repulsive Coulomb interactions, it is immediately clear that fluctuations of all Wishart eigenvalues towards large values are extremely rare. Due to the repulsive interactions, a fluctuation of the smallest eigenvalue to scales of order the typical eigenvalue, $N \sigma^2$, corresponds to a configuration with free energy $H\sim  N^2/2$, and is therefore super-exponentially suppressed, with probability $\sim e^{-N^2}$. The precise probability density function for the smallest eigenvalue is known analytically in terms of hypergeometric functions; for $\beta=1$ and $N \gg 1$ one  finds \cite{Edelman:1988:ECN:58846.58854,2010arXiv1005.4515M}
\be\label{rhomin}
\rho_{\lambda_{{\rm{min}}}}(\lambda)={1\over 2\sigma^2} \left(\sqrt{N \sigma^2\over \lambda}+N\right)\exp\left(-\sqrt{N\lambda\over \sigma^2} -{N\lambda\over 2\sigma^2} \right)\,.
\ee
It follows that the median  size of the smallest eigenvalue is $\tilde{ \lambda}_{{\rm{min}}}={\calc}  \sigma^2/N$, where
\bea
\label{cmedian}
\calc=2+\log(4)-2\sqrt{1+\log(4)}\approx 0.30\,.
\eea
Using (\ref{rhomin}) we immediately have the probability distribution of the inverse of the smallest eigenvalue:\footnote{The probability distribution of the inverse of a random variable with distribution $\rho(\lambda)$ is given by $\rho_{-1}(\mu)=1/\mu^2 \rho(1/\mu)$.}
\be
\rho(\lambda)^{{\rm{Wis}}_{1/{\rm{min}}}}={1\over 2(\lambda\sigma^2)^{3/2}}\left( \sqrt{N}+{N\over \sqrt{\lambda\sigma^2}}\right)\exp\left(-\sqrt{N\over\sigma^2\lambda} -{N\over2\sigma^2\lambda} \right) \,.
\ee
This scales as $1/\lambda^{3/2}$ for large $\lambda$,  so the distribution is heavy-tailed.
On the other hand, a fluctuation to small inverse eigenvalues is heavily suppressed. Therefore, the smallest eigenvalue can easily be much smaller than its typical value, but not much larger.   (In  the context of our analysis of the diameters of axion fundamental domains, this fact  about the smallest eigenvalue of a Wishart matrix implies that the diameter can significantly exceed the lower bounds derived in this work.)

Another useful property follows from the rotational symmetries inherent to the Wigner and Wishart ensembles: eigenvector delocalization (cf.~\cite{TaoVu}).
Given a  collection of matrices drawn from the Wishart ensemble, the eigenvectors are  with high probability uniformly distributed on the sphere $S^{N-1}$.
As a result,  the entries of normalized eigenvectors are are normally distributed with vanishing mean and variance $1/N$. The median size of the the largest-magnitude entry of a delocalized eigenvector immediately evaluates to
\be\label{maxcomp}
\text{Max}(\{|\psi_{i}|\})={\sqrt{2} \,\text{erf}^{-1}(2^{-1/N})\over \sqrt{N}} \equiv {\ell_N \over \sqrt{N}}\,,
\ee
Thus, up to logarithmic corrections, which we capture in the factor $\ell_N$, the largest entry of the eigenvector is given by $1/\sqrt{N}$ \cite{Rudelson1}.

This result also holds for the inverse of a rotationally symmetric matrix, because the eigenvectors of a diagonalizable matrix are unaffected by taking the inverse. Furthermore, one can verify numerically that matrices of the form
\be\label{abc}
\mathbf A=\mathbf B^\top \mathbf C \mathbf B
\ee
obey eigenvector delocalization, as long as the matrices $\mathbf B^\top \mathbf B$ form a rotationally invariant ensemble, independent of the properties of $\mathbf C$. We will encounter potentially non-rotationally invariant matrices $\mathbf C$, appearing in the form (\ref{abc}), where eigenvector delocalization of $\mathbf A$ still holds.

\subsection{Approach to universality}
It is important to address the conditions under which random matrices approach the universal regime. In particular, the classical random matrix ensembles are defined in terms of non-heavy tailed entries, i.e.~the cumulative distribution function of the entries decays at least exponentially.
However, there are more general matrices that still approach universality.

Let us consider the example of a $N\times N$ unit matrix that is perturbed by a matrix $\boldsymbol \delta \mathbf Q$, where  $\boldsymbol \delta \mathbf Q$ is a random matrix with real i.i.d.~elements with  the Gaussian distribution  ${\mathcal N}(0,\sigma_{\delta Q})$:
\be
\mathbf Q={\mathbb 1}+\boldsymbol \delta \mathbf Q\,.
\ee
The matrix $\mathbf Q^\top \mathbf Q$ is given by
\be
\mathbf Q^\top \mathbf Q={\mathbb 1}+(\boldsymbol \delta \mathbf Q_{\sigma_{\delta Q}} +\boldsymbol \delta \mathbf Q_{\sigma_{\delta Q}}^\top)+\boldsymbol \delta \mathbf Q_{\sigma_{\delta Q}}^\top \boldsymbol \delta \mathbf Q_{\sigma_{\delta Q}}\,,
\ee
in which the first term has eigenvalues of order $1$, the second term is a Wigner matrix with eigenvalues of order $\sqrt{8 N\sigma_{\delta Q}^2}$, and the third term, which is a Wishart matrix, has eigenvalues of order $4\sigma_{\delta Q}^2 N$. The matrix $\mathbf Q^\top \mathbf Q$ is well approximated by a Wishart matrix for $\sigma_{\delta Q}\gtrsim {1/(2\sqrt{N})}$. While this parametric scaling is confirmed by numerical studies, we observe that the smallest eigenvalue actually approaches the Wishart result for $\sigma_{\delta Q}\gtrsim \sqrt{2}/\sqrt{N}$.

Let us consider the example of $\mathbf Q={\mathbb 1}+\boldsymbol \delta \mathbf Q$, where $\boldsymbol \delta \mathbf Q$ consists of a matrix with $N_{\delta Q}$ random entries equal to one, with all other entries vanishing. The quadratic norm of the entries of $\boldsymbol \delta \mathbf Q$ evaluates to (for $N_{\delta Q}\ll N^2$)
\be
\sigma^2_{\delta Q}=N_{\delta Q}/N^2\,.
\ee

As we noted above, the matrix $\mathbf Q^\top \mathbf Q$ approaches universality for $\sigma_{\delta Q}^2\gtrsim {2\over  N}$, which is satisfied for
\be\label{minevuniversal}
N_{\delta Q}\gtrsim {2N}\,.
\ee
Therefore, we have a lower limit for $\sigma_{\delta Q}$ (assuming only unit entries of $\mathbf Q$),
\be
\sigma_{\delta Q}\gtrsim {\sqrt{2\over N}}\,,
\ee
which corresponds to $N_{\delta Q}=2N$. Thus, the universal regime is approached by perturbing a unit $N\times N$ matrix by $2N$ random elements.

\section{$\mathbf Q$ in Calabi-Yau Hypersurfaces in Toric Varieties}\label{app:toric}

In this appendix we  briefly explore the form of $\mathbf q$ in two examples of Calabi-Yau hypersurfaces in toric varieties.  We will simply demonstrate the nontriviality of certain $\mathbf q$, and not concern ourselves with explicit orientifold involutions, etc.
All reflexive polytopes in four dimensions are available in the Kreuzer-Skarke database~\cite{KS_database}. Triangulation of the corresponding polytope yields a simplicial toric variety with at most pointlike singularities~\cite{Batyrev}, which are missed by a generic Calabi-Yau hypersurface. We use the algorithm presented in the appendix of~\cite{Long:2014fba} to triangulate the polytopes and define the toric variety.
For each ray $v_i$ in the fan that defines the toric variety there is a corresponding homogeneous coordinate $x_0$, the vanishing of which defines a divisor $D_i$. The $D_i$ are called the toric divisors, and define irreducible hypersurfaces in the toric variety. In the following we will refer to both the divisor and its cohomological dual as $D_i$. A subset of the $D_i$ form a basis for $H^{1,1}(X_3 , \mathbb{Z})$. To see which linear combinations of $D_i$ contribute to the nonperturbative superpotential we need to compute certain Hodge numbers of the toric divisors. This can be done using the program \textbf{cohomCalg}~\cite{Blumenhagen:2010ed}, an implementation of the algorithm suggested and proved in~\cite{Blumenhagen:2010pv, Rahn:2010fm, 2011JMP....52c3506J}, which uses the Koszul sequence to calculate line bundle topology in toric varieties. We then calculate the leading order contributions to the nonperturbative superpotential, which defines the $\mathbf q$ matrix.
As a first example, we consider the Calabi-Yau hypersurface in the toric variety given in Table~\ref{table:toric1}, which we denote by $V_A$.

\begin{table*}
\ra{1.3}
\begin{center}
\begin{tabular}{r r r r r r r r r r  r c r }\toprule
\label{table:toric1}
$x_0$&$x_1$& $x_2$& $x_3$& $x_4$& $x_5$& $x_6$& $x_7$& $x_8$& $x_9$& $x_{10}$ &\phantom{a}&$\phantom{i}p$  \\
\cmidrule{1-11}\cmidrule{13-13}
$-1$&$0$& $0$& $1$& $0$& $-2$& $0$& $1$& $0$& $0$& $1$ &&$0$ \\
$0$&$0$& $0$& $0$& $-1$& $0$& $0$& $1$& $1$& $-1$& $0$ &&$0$  \\
$0$&$0$& $0$& $0$& $0$& $1$& $1$& $0$& $0$& $0$& $-2$ &&$0$  \\
$0$&$0$& $0$& $0$& $1$& $0$& $-1$& $0$& $0$& $0$& $1$ & &$1$\\
$0$&$0$& $1$& $0$& $0$& $0$& $0$& $0$& $-2$& $1$& $0$ &&$0$ \\
$0$&$1$& $0$& $-1$& $0$& $2$& $0$& $1$& $0$& $-1$& $0$ &&$2$ \\
$0$&$1$& $0$& $0$& $-1$& $0$& $1$& $0$& $0$& $0$& $0$ & &$0$\\
\bottomrule
\end{tabular}\label{rmt}
\caption{Charges for $V_A$.}
\end{center}
\end{table*}
The Stanley-Reisner ideal is given by
\begin{eqnarray}
\text{SR} =\{x_1 x_6, x_1 x_{10} x_3 x_8, x_0 x_9 x_2 x_9, x_2 x_7, x_7 x_8, x_0 x_2, x_0 x_8,x_4 x_5, x_5 x_6, x_{10} x_4, \nonumber \\ x_1 x_5 x_6,
x_3 x_4 x_9, x_3 x_6 x_9,  x_3 x_9 x_{10}, x_3 x_4 x_7, x_3 x_6 x_7, x_3 x_7 x_{10}, x_0 x_1 x_5\}\,.
\end{eqnarray}
This toric variety defines a Calabi-Yau hypersurface with $h^{1,1} = 7$.  We take $D_i = \{ D_{10},D_9, D_8, D_7,D_6,D_5,D_4\}$ as a basis for divisors. The divisors $\{D_{10},D_9,D_6,D_5,D_4\}$ are rigid toric divisors.  Moreover, the combinations $D_9 + D_7$, $D_9 + D_8$  are rigid. The ${\mathbf q}$ matrix is then given by

\[ \mathbf q = \begin{blockarray}{cccccccc}
& D_{10} & D_9 & D_8 & D_7 & D_6 & D_5 & D_4 \\
 \begin{block}{c(ccccccc)}
  W_1 & 1 & 0 & 0 & 0 & 0 & 0 & 0 \\
  W_2 & 0 & 1 & 0 & 0 & 0 & 0 & 0  \\
  W_3 & 0 & 1 & 1 & 0 & 0 & 0 & 0 \\
  W_4 & 0 & 1 & 0 & 1 & 0 & 0 & 1 \\
  W_5 & 0 & 0 & 0 & 0 & 1 & 0 & 0 \\
  W_6 & 0 & 0 & 0 & 0 & 0 & 1 & 0 \\
  W_7 & 0 & 0 & 0 & 0 & 0 & 0 & 1 \\
  \end{block}
  \end{blockarray}\,\,.
  \]
Here each $W_i$, $i = 1 \dots 7$, denotes the $i$th contribution to the nonperturbative superpotential. We have kept only the leading contributions to the nonperturbative superpotential, neglecting higher-order terms, e.g.~from the rigid cycle $D_{10} + D_9$.

The K\"{a}hler cone conditions present a difficulty in this example.  If we demand that each of the holomorphic curves, given by generators of the Mori cone, has area of at least $1$, then the four-cycles that appear in the superpotential are forced to become very large.
As a result,  the nonperturbative superpotential --- and correspondingly, the scalar potential --- become extremely small in Planck units, precluding moduli stabilization near the GUT scale.
For the purpose of constructing models of large-field inflation, one would like to find Calabi-Yau manifolds with ``mild'' topology, by which we mean that the divisor volumes do not grow rapidly with the curve areas. The DDFGK compactification described in \S\ref{sec:DDFGK} is one such example, but it would be valuable to characterize this issue more generally.

As a second example, we consider the Calabi-Yau hypersurface in the toric variety in Table~\ref{table:toric2}, denoted $V_B$.
\begin{table*}
\ra{1.3}
\begin{center}
\begin{tabular}{r r r r r r r r r r  r c r }\toprule
\label{table:toric2}
$x_0$&$x_1$& $x_2$& $x_3$& $x_4$& $x_5$& $x_6$& $x_7$& $x_8$& $x_9$& $x_{10}$ &\phantom{a}&$\phantom{i}p$  \\
\cmidrule{1-11}\cmidrule{13-13}
$-1$&$0$& $1$& $0$& $-1$& $0$& $0$& $1$& $0$& $0$& $1$ &&$1$ \\
$-1$&$1$& $1$& $0$& $0$& $3$& $0$& $0$& $-1$& $0$& $0$ &&$3$  \\
$0$&$0$& $-1$& $-1$& $1$& $0$& $0$& $0$& $1$& $0$& $0$ &&$0$  \\
$0$&$0$& $0$& $0$& $0$& $0$& $-2$& $0$& $0$& $1$& $1$  &&$0$\\
$0$&$0$& $0$& $1$& $0$& $-1$& $1$& $0$& $0$& $-1$& $0$ &&$0$ \\
$0$&$0$& $1$& $-1$& $-1$& $0$& $0$& $0$& $0$& $1$& $0$ &&$0$ \\
$0$&$1$& $1$& $1$& $0$& $0$& $0$& $-2$& $-1$& $0$& $0$  &&$0$\\
\bottomrule
\end{tabular}\label{rmt}
\caption{Charges for $V_B$.}
\end{center}
\end{table*}
The Stanley-Reisner ideal is given by
\begin{eqnarray}
\text{SR} =\{x_{4}x_{8},x_{6}x_{8},x_{8}x_{9},x_{10}x_{8},x_{2}x_{6},x_{2}x_{9},x_{0}x_{3},x_{3}x_{6},x_{10}x_{3},x_{5}x_{7},x_{0}x_{9}, \nonumber \\ x_{10}x_{9},x_{1}x_{4},x_{0}x_{6},x_{1}x_{2}x_{3},x_{1}x_{2}x_{5},x_{0}x_{4}x_{5},x_{1}x_{10}x_{2},x_{10}x_{2}x_{7},x_{0}x_{7}x_{8}\}\,.
\end{eqnarray}
This toric variety defines a Calabi-Yau hypersurface with $h^{1,1} = 7$.  We take $D_i = \{ D_{10},D_9, D_8, D_7,D_6,D_5,D_4\}$ as a basis for divisors. The divisors $\{D_{10},D_9,D_8,D_5,D_4\}$ are rigid toric divisors, while $\{D_7,D_6\}$ are exact Wilson divisors with $h^{0,1} = 1$. Moreover, the combinations $D_6 + D_9$, $D_4 + D_7$, and $D_{10} + D_6$ are rigid. The ${\mathbf q}$ matrix is then given by

\[ \mathbf q = \begin{blockarray}{cccccccc}
& D_{10} & D_9 & D_8 & D_7 & D_6 & D_5 & D_4 \\
 \begin{block}{c(ccccccc)}
  W_1 & 1 & 0 & 0 & 0 & 0 & 0 & 0 \\
  W_2 & 0 & 1 & 0 & 0 & 0 & 0 & 0  \\
  W_3 & 0 & 0 & 1 & 0 & 0 & 0 & 0 \\
  W_4 & 0 & 0 & 0 & 1 & 0 & 0 & 1 \\
  W_5 & 0 & 1 & 0 & 0 & 1 & 0 & 0 \\
  W_6 & 0 & 0 & 0 & 0 & 0 & 1 & 0 \\
  W_7 & 0 & 0 & 0 & 0 & 0 & 0 & 1 \\
  W_8 & 1 & 0 & 0 & 0 & 1 & 0 & 0 \\
  \end{block}
  \end{blockarray}\,\,.
  \]

Again each $W_i$, $i = 1 \dots 8$, denotes the $i$th contribution to the nonperturbative superpotential. Note that at leading order there are eight contribution to $W$ for seven divisors.

To build a vacuum of string theory in which inflation can occur, one must take into account many more details, such as a consistent orientifold with tadpole cancellation and moduli stabilization. Here we  have simply demonstrated the nontriviality of $\mathbf q$ at moderate $h^{1,1}$.  A statistical study of the form of $\mathbf q$ at moderate to large $h^{1,1}$ would be an interesting direction for the future.

\bibliographystyle{modifiedJHEP}
\bibliography{refs}
\end{document}